\documentstyle[12pt,rotate]{article}

\input epsf.sty

\newcommand{\ft}[2]{{\textstyle\frac{#1}{#2}}}

\newcommand{\g}{{\sl g}}
\newcommand{\BRST}{{\rm BRST}}
\newcommand{\cA}{{\cal A}}
\newcommand{\cJ}{{\cal J}}

\newcommand{\cX}{{\cal X}}
\newcommand{\cD}{{\cal D}}
\newcommand{\cK}{{\cal K}}

\newcommand{\cV}{{\cal V}}

\newcommand{\cO}{{\cal O}}

\newcommand{\cP}{{\cal P}}
\newcommand{\cN}{{\cal N}}
\newcommand{\cR}{{\cal R}}
\newcommand{\cQ}{{\cal Q}}
\newcommand{\cF}{{\cal F}}

\newcommand{\cS}{{\cal S}}
\newcommand{\cT}{{\cal T}}

\newcommand{\cL}{{\cal L}}
\newcommand{\cU}{{\cal U}}

\font\cmss=cmss12 
\def\1{\hbox{{1}\kern-.25em\hbox{l}}}
\def\bfZ{\relax{\hbox{\cmss Z\kern-.4em Z}}}

\begin{document}
\begin{titlepage}

\centerline{\large \bf Superconformal constraints for QCD conformal
                       anomalies. }

\vspace{15mm}

\centerline{\bf A.V. Belitsky$^a$, D. M\"uller$^b$}

\vspace{15mm}

\centerline{\it $^a$C.N.\ Yang Institute for Theoretical Physics}
\centerline{\it State University of New York at Stony Brook}
\centerline{\it NY 11794-3840, Stony Brook, USA}

\vspace{5mm}

\centerline{\it $^b$Institut f\"ur Theoretische Physik, Universit\"at
                Regensburg}
\centerline{\it D-93040 Regensburg, Germany}

\vspace{20mm}

\centerline{\bf Abstract}

\hspace{0.5cm}

\noindent Anomalous superconformal Ward identities and commutator algebra in
$\cN = 1$ super-Yang-Mills theory give rise to constraints between the QCD
special conformal anomalies of conformal composite operators. We evaluate
the superconformal anomalies that appear in the product of renormalized
conformal operators and the trace anomaly in the supersymmetric spinor
current and check the constraints at one-loop order. In this way we prove
the universality of QCD conformal anomalies, which define the non-diagonal
part of the anomalous dimension matrix responsible for scaling violations of
exclusive QCD amplitudes at the next-to-leading order.

\vspace{5cm}


\vspace{0.5cm}


\end{titlepage}

\section{Introduction.}

Supersymmetric models \cite{Son85}, having higher space-time symmetry as
compared to conventional ones, provide a strong consistency requirement
on theoretical predictions. For the purposes of testing massless QCD
calculations an especially illuminating example is $\cN = 1$ supersymmetric
Yang-Mills theory, since both models have, up to a difference in colour
representation of fermion fields, the same Lagrangian. Thus, we can map
a QCD result to an $\cN = 1$ super-Yang-Mills theory one by identifying
the colour Casimir operators in corresponding representations, i.e.\
$C_A = C_F = 2 N_f T_F \equiv N_c$. After this procedure a QCD result
has to fulfill constraints arising from supersymmetry which connects
gluonic and quark sectors of the theory. In this way the use of
supersymmetry has allowed to find a set of identities
\cite{BukFroKurLip85,BelMueSch98} between the entries of the forward
anomalous dimension matrices of leading twist-two composite operators.
They were valuable to clarify subtleties appearing in two-loop
computations of anomalous dimensions. On tree level both theories are
invariant under conformal transformations. Thus, the $\cN = 1$
supersymmetric Yang-Mills theory is also invariant under superconformal
transformations \cite{WesZum70,Fer74}, which can give rise to a new set
of constraints for certain conformal quantities which appear in the
special conformal Ward identities for composite operators, the so-called
special conformal matrix. Besides the breaking of conformal symmetry on
quantum level by the trace anomaly in the energy-momentum tensor, we
also have to deal with a superconformal anomaly due to the non-vanishing
trace in a supersymmetric spinor current. Nevertheless, an explicit
calculation of this anomaly will allow to check the special conformal
anomalies calculated in QCD.

Composite operators appear in various QCD applications by means of the
operator product expansion and consequently their hadronic matrix elements
contain a non-perturbative input, which is needed as an initial condition
for the solution of the evolution equations. In the case of exclusive
processes the off-forwardness of hadronic matrix elements, given in terms
of distribution amplitudes and skewed parton distributions, requires
operators with total derivatives. To ensure that the twist-two operators
do not mix under renormalization at leading order in coupling constant,
i.e.\ their anomalous dimension matrix has the diagonal form
$\mbox{\boldmath$\gamma$}_j \delta_{jk}$, it is necessary to arrange
the operators in such a way that they have a covariant behaviour
under conformal transformations. This can be easily done. However,
beyond one-loop approximation the anomalous dimension matrix
$\mbox{\boldmath$\gamma$}_{jk} $ develops non-zero non-diagonal, $j > k$,
elements $\mbox{\boldmath$\gamma$}^{\rm ND}_{jk} \propto \cO (\alpha_s^2)$.

The ordinary conformal algebra provides severe restrictions
\cite{Mue94,BelMul98b} on the non-forward anomalous dimensions
$\mbox{\boldmath$\gamma$}$ of the conformal operators. In Refs.\
\cite{Mue94,BelMul98b} we have developed a formalism based on the use of the
broken conformal Ward identities for evaluation of the non-diagonal part,
$\mbox{\boldmath$\gamma$}^{\rm ND}$, of the complete anomalous dimensions
matrix $\mbox{\boldmath$\gamma$} = \mbox{\boldmath$\gamma$}^{\rm D} +
\mbox{\boldmath$\gamma$}^{\rm ND}$. This non-diagonal part arises entirely
due to the violation of the special conformal symmetry in perturbation theory.
The corresponding anomalies have been calculated to one-loop order accuracy
in the minimal subtraction scheme using dimensional regularization, which
imply the two-loop approximation for $\mbox{\boldmath$\gamma$}^{\rm ND}$.
To check our results, one can employ $\cN = 1$ super Yang-Mills constraints,
valid in a renormalization scheme that respects supersymmetry, for the
entries of the non-forward anomalous dimension matrix, derived in
\cite{BelMueSch98}. Unfortunately, this is not the case for the dimensionally
regularized theory. Thus, one has to find finite renormalization constants 
from the latter to the dimensional reduction scheme, which is
expected to preserve the supersymmetry. But there arises a subtlety in the
evaluation of this rotation matrix for the gluonic sector\footnote{This
complication does not show up in the forward kinematics.} which prevents it
to be unambiguously fixed \cite{BelMueSch98}. Nevertheless, our result for
two-loop non-forward anomalous dimensions is supported by the fact that
the constraints can be fulfilled by a finite multiplicative renormalization,
which proves the existence of a supersymmetric regularization scheme.

Alternatively, we derive in this paper constraints directly for the special
conformal anomalies at one-loop level and show that they are indeed satisfied.
Our consequent presentation is organized as follows. In section
\ref{Sec-Prel} we define conformal operators, their anomalous dimensions,
and relations of the latter to conformal anomalies. Section
\ref{LargangAnomalies} is devoted to the study of translational and
conformal super-anomalies on Lagrangian level in the dimensional
regularization scheme. Then in section \ref{WardIdentitesSC} we present
transformation properties of conformal operators under relevant
superconformal variations required for a derivation of the Ward identities
discussed in the same section. In section \ref{ConstraintsSUSY} we give
a derivation of relations between the scale and special conformal
anomalies of conformal operators. Furthermore, we show that the latter
acquire anomalous contributions originating from the product of the trace
anomaly in the spinor current and conformal operators. They are
explicitly evaluated in section \ref{ExplicitCalcAnom}, where it is
demonstrated that indeed the anomalous constraints are fulfilled with
special conformal anomalies from \cite{BelMul98b}. Finally, we conclude.
A few appendices are devoted to technical details that we found
inappropriate to include in the body.

\section{Preliminaries.}
\label{Sec-Prel}

In this paper we discuss relations between the QCD scale and special
conformal anomalies of conformal operators implied by the $\cN = 1$
supersymmetry. In $\cN = 1$ super Yang-Mills theory we introduce the
conformal operators (for chiral even sector discussed throughout)
\begin{equation}
\label{treeCO}
{^Q\!\cO^i_{jl}}
= \frac{1}{2}
\bar\psi^a_+ (i \partial_+)^l
C^{3/2}_j
\left( \frac{\stackrel{\leftrightarrow}{\cD}_+}{\partial_+} \right)
{\mit\Gamma}^i
\psi^a_+ , \quad
{^G\!\cO^i_{jl}}
=
G^{a \perp}_{+ \mu} (i \partial_+)^{l-1}
C^{5/2}_{j - 1}
\left(
\frac{\stackrel{\leftrightarrow}{\cD}_+}{\partial_+}
\right)
\cT^i_{\mu\nu}
G^{a \perp}_{\nu +} ,
\end{equation}
where $C^{\nu}_j$ are the Gegenbauer polynomials and the tensor structures
are ${\mit\Gamma}^{(V; A)} = (\gamma_+; \gamma_+\gamma_5)$, $\cT^{(V;
A)}_{\mu\nu} = (g^\perp_{\mu\nu} \equiv g_{\mu\nu} - n_\mu n^\star_\nu -
n^\star_\mu n_\nu; i \epsilon_{\mu\nu\rho\sigma} n^\star_\rho n_\sigma)$. We
use the convention $\stackrel{\phantom{\rightarrow}}{\partial} \, = \,
\stackrel{\rightarrow}{\partial} \!\!+\!\! \stackrel{\leftarrow}{\partial}$
and $\stackrel{\leftrightarrow}{\cD} \, = \, \stackrel{\rightarrow}{\cD}\!
-\! \stackrel{\leftarrow}{\cD}$ with adjoint covariant derivative defined by
$\cD_\mu^{ab} = \partial_\mu \delta^{ab} + \g f^{acb} B_\mu^c$. The
`$+$'-sign as a subscript stands for contraction with the light-like vector
$n_\mu$ which specifies a direction along the light cone. For the latter
purposes we introduce another vector $n^\star_\mu$ such as $n^2 = n^{\star
2} = 0$ and $n \cdot n^\star = 1$. Obviously, the only difference from QCD
arises in the gluino, which we loosely call quark, sector, which now belongs
to the adjoint representation of the colour group. The factor $\ft12$ in
Eq.\ (\ref{treeCO}) is related to the Majorana nature of the quarks in the
model.

The renormalization group equation for these operators looks like
\begin{equation}
\frac{d}{d \ln \mu} [\mbox{\boldmath$\cO$}_{jl}]
= - \sum_{k = 0}^{j} \mbox{\boldmath$\gamma$}_{jk}
[\mbox{\boldmath$\cO$}_{kl}],
\end{equation}
where the square brackets will denote the renormalized operators defined by
$[\mbox{\boldmath$\cO$}_{jl}] = \sum_{k = 0}^{j} \mbox{\boldmath$Z$}_{jk}
\mbox{\boldmath$\cO$}_{k l}$, with renormalization constant matrix
$\mbox{\boldmath$Z$}_{jk}$, which generate finite Green functions with
elementary field operators $\phi = \{ \psi, B_\mu \}$. We use everywhere the
matrix notation and introduce the vector $\mbox{\boldmath$\cO$} = \left( {
{^Q\!\cO} \atop {^G\!\cO}} \right)$ of quark and gluon conformal operators,
which mix with each other under renormalization. As mentioned in the
introduction, the only modification of a given QCD result is to identify the
colour Casimir operators. Since at leading order the conformal anomalies
have a unique colour structure\footnote{More precisely, the anomalous
dimensions in the gluon-gluon channel have in addition to the $C_A$ term
also trivial $N_f$ dependent contributions, which arise from the
self-energy insertion.} it presents no difficulty to disentangle the
separate components.

We introduce as well the fermionic operator which is related to the bosonic
ones (\ref{treeCO}) by supersymmetry
\begin{equation}
\label{FermConfOp}
{^F\!\cO^i_{jl}}
= G^{a \perp}_{+ \mu} (i \partial_+)^l
P^{(2,1)}_j\!\! \left(
\frac{\stackrel{\leftrightarrow}{\cD}_+}{\partial_+}
\right)
\cF^i_\mu
\psi^a_+ .
\end{equation}
Here $P_j^{(a,b)}$ are the Jacobi polynomials and the vertices read 
$\cF^{(V; A)} = (\gamma^\perp_\mu; \gamma^\perp_\mu \gamma_5)$. The 
operators that form a representation of the supersymmetry algebra are 
defined by linear combinations of (\ref{treeCO})
\begin{eqnarray}
\left\{
\begin{array}{c}
\cS^a \\
\cP^a
\end{array}
\right\}_{jl}
=
{^Q\omega^a_j} \
{^Q\! \cO}^{\mit\Gamma}_{jl}
+
{^G\omega^a_j} \
{^G\! \cO}^{\mit\Gamma}_{jl},
\qquad
\cV_{jl} = \varrho_j {^F\!\cO^V} ,
\qquad
\cU_{jl} = \varrho_j {^F\!\cO^A} ,
\end{eqnarray}
with ${\mit\Gamma} = V(A)$ standing for the $\cS$ ($\cP$) operator,
and coefficients ${^Q\omega^1_j} = 1$, ${^G\omega^1_j} = \frac{6}{j}$,
${^Q\omega^2_j} = - \frac{j + 3}{j + 1}$, ${^G\omega^2_j}
= \frac{6}{j + 1}$ and $\varrho_j = \frac{(j + 2)(j + 3)}{(j + 1)}$.
Obviously, $\cU = - \gamma_5 \cV$. Note that the bosonic and fermionic
conformal operators form the $\cN = 1$ chiral superfield
\begin{equation}
{\mit\Phi} = \cA + 2 \theta \chi - \theta^2 \cF ,
\end{equation}
with operators $\cS^1_{jl}$ and $\cP^1_{jl}$ ($\cS^2_{jl}$ and
$\cP^2_{jl}$) being the real and imaginary parts of the $\cA$
($\cF^\dagger$) complex fields, and $\cV_{j - 1, l}$ identified with the
Majorana fermion $\left( { \chi^\alpha \atop \bar\chi_{\dot\alpha}}
\right)$ constructed from the Weyl spinor $\chi$. Transformation between
operators under supersymmetry arises from the conventional equation
$[\bar\zeta \cQ, {\mit\Phi}]_- = [\zeta \cQ + \bar\cQ \bar\zeta,
{\mit\Phi}]_- = \left\{ \zeta r + \bar r \bar\zeta \right\} {\mit\Phi}$,
with $r = i \frac{\partial}{\partial \theta}$ and $\bar r = - i
\frac{\partial}{\partial\bar\theta} + 2 \theta \not\!\partial$.

Now let us shortly point out, how the non-diagonal part of the anomalous
dimension matrix is induced by the special conformal anomaly matrix. In four
dimensional space-time the fifteen-parameter conformal group $SO(4,2)$ is
defined by its algebra containing the Poincar\'e, dilatation $\cD$ and
special conformal $\cK_\mu$ generators. The conformal anomalies are
defined by the renormalized Ward identities. The generic form of the
latter, however, written in an unrenormalized cast, reads
\begin{equation}
\label{WardIdentity}
\langle [\mbox{\boldmath$\cO$}_{jl}] \delta \cX \rangle
= - \langle \delta [\mbox{\boldmath$\cO$}_{jl}] \cX \rangle
- \langle i [\mbox{\boldmath$\cO$}_{jl}] \delta S \cX \rangle ,
\end{equation}
where $\cX = \prod_{\ell} \phi (x_\ell)$ is a product of elementary fields
appearing in the classical Lagrangian. Here $\delta$ is any of the
variations from the symmetry algebra in question. When the transformation is
a symmetry of the theory on quantum level then $\delta S = 0$ up to possible
BRST exact operators. In the (dimensionally) regularized theory the action
does not vanish anymore for conformal, i.e.\  both scaling, $\delta^S
\phi(x)= i [\phi(x),\cD] = -\left(d_\phi +x_\nu \partial_\nu \right) \phi
(x)$, and special conformal, $\delta^C_\mu \phi(x)= i [\phi(x),\cK_\mu] =
-\left( 2d_\phi x_\mu - x^2 \partial_\mu + 2 x_\mu x_\nu \partial_\nu - 2i
x_\nu \Sigma_{\mu\nu} \right) \phi (x)$, variations, where $d_\phi$ and
$\Sigma_{\mu\nu}$ are the canonical dimension and the spin operator of the
field $\phi$, respectively. Thus, the renormalization of the operator
product $i [\mbox{\boldmath$\cO$}_{jl}] \delta S $ is responsible for the
conformal anomalies.

Moreover, the commutator $[\cD, \cK_-]_- = i\cK_-$, where $\cK_-$ being the
$n^\star_\mu$-light-cone projection of $\cK_\mu$, provides a connection
between the conformal anomalies. In Ref.\ \cite{BelMul98b} the non-diagonal
elements of the next-to-leading anomalous dimensions
\begin{equation}
\label{PreAnoDim}
\mbox{\boldmath$\gamma$}^{{\rm ND}(1)} \!
=
[ \mbox{\boldmath$\gamma$}^{{\rm D}(0)}, \mbox{\boldmath$d$}\,
( \beta_0 - \mbox{\boldmath$\gamma$}^{{\rm D}(0)} )
+ \mbox{\boldmath$g$} ]_-
\end{equation}
of QCD quark and gluon conformal operators
were found in terms of one loop special conformal anomaly matrix
\begin{equation}
\label{SpecialConformalAnomaly}
a_{jk}^{- 1} (B) \mbox{\boldmath$\gamma$}^{c(0)}_{jk}
\equiv
- \mbox{\boldmath$d$}_{jk}
\left( \mbox{\boldmath$\gamma$}^{{\rm D}(0)}_k
- \beta_0 \mbox{\boldmath$P$}^G \right)
+ \mbox{\boldmath$g$}_{jk} .
\end{equation}
It is constructed out of the leading order anomalous dimensions of
conformal operators $\mbox{\boldmath$\gamma$}^{{\rm D}(0)}$ and the
first expansion coefficient of the QCD $\beta$-function $\beta_0 =
\frac{4}{3} T_F N_f - \frac{11}{3} C_A$ times $\left. a_{jk} (B)
d_{jk} (F) \right|_{j>k} = - 2 (2k + 3)$ with $a_{jk}$
matrix from the conformal transformation of $\mbox{\boldmath$\cO$}_{jk}$
(see Eq.\ (\ref{ConformalVariation}) below). The projector
$\mbox{\boldmath$P$}^G = \left( { 0 \ 0 \atop 0 \ 1}\right)$ in Eq.\
(\ref{SpecialConformalAnomaly}) singles out the gluonic component.
Finally, the $\mbox{\boldmath$g$}$-matrix has appeared from the
renormalization of the product of the conformal operator
$[\mbox{\boldmath$\cO$}_{jk}]$ and the integrated trace anomaly
$\delta^C_- S \propto - \int d^d x 2 x_- {\mit\Theta}_{\mu\mu} (x)$
in the energy-momentum tensor
\begin{equation}
\label{QGspecialConfAnom}
[\mbox{\boldmath$\cO$}_{jl}] \delta^C_- S
= i \sum_{k = 0}^{j}
\mbox{\boldmath$\gamma$}^c_{jk} [\mbox{\boldmath$\cO$}_{k l - 1}] + \cdots.
\end{equation}
In the dimensionally regularized theory, i.e.\ $d = 4 - 2\varepsilon $,
the conformal variations of the QCD action can be calculated in a
straightforward manner. Choosing the scaling dimensions of the physical
fields equal to their canonical values in four dimensions\footnote{This
choice is legitimate since the infinitesimal conformal variation is linear
in $d_\phi$ and thus does not affect the Ward identities, since the
anomalous part will show up as a renormalization counterterm of the
product of conformal and equation-of-motion operators.} ($d_\psi = \ft32$,
$d_B = 1$) and setting the scaling dimensions of the ghost fields as
$d_\omega = 0$ and $d_{\bar\omega} = d - 2$, the result
\begin{eqnarray}
\label{ConformalVarS}
\delta^B S = \int d^d x\ w^B (x) \Bigg\{
\!\!\!&-&\!\!\! \frac{d - 4}{2}
\left( {\cal O}_A (x) + {\cal O}_B (x) + {\mit \Omega}_{\bar \omega} (x)
- {\mit \Omega}_{\bar\psi\psi} (x) - {\mit \Omega}_D (x) \right)
\nonumber\\
&+&\!\!\! (d - 2)\,\partial_\mu {\cal O}_{B\mu}(x)
\Bigg\},
\qquad\mbox{with}\qquad
B = \{D, C\},
\end{eqnarray}
is decomposed in operators that can be easily classified according to
their renormalization properties. Here the weight function reads $w^D = 1$
and $w^C = 2 x_-$ for scale and special conformal transformations,
respectively. We introduced the following set of type A and B operators
\begin{eqnarray}
{\cal O}_A (x) = \frac{1}{2} \left( G^a_{\mu\nu} \right)^2, \ \
{\cal O}_B (x) = \delta^{\rm BRST}
\left( \bar{\omega}^a\partial_\mu B_\mu^a \right), \ \
{\cal O}_{B\mu} (x) = \delta^{\rm BRST}
\left( \bar{\omega}^a B^a_\mu \right),
\end{eqnarray}
as well as class C equation-of-motion operators
\begin{equation}
{\mit \Omega}_G (x) = B^a_\mu \frac{\delta S}{\delta B^a_\mu},
\quad
{\mit \Omega}_{\bar\psi\psi} (x)
= \frac{\delta S}{\delta \psi} \psi
+ \bar\psi \frac{\delta S}{\delta \bar\psi},
\quad
{\mit \Omega}_{\bar \omega} (x)
= \bar \omega^a \frac{\delta S}{\delta \bar \omega^a} ,
\quad
{\mit \Omega}_D (x) = D^a \frac{\delta S}{\delta D^a}.
\end{equation}
The renormalization of Eq.\ (\ref{ConformalVarS}) is straightforward
and the renormalization of the operator products and the resulting
renormalized conformal Ward identities are given in Ref.\ \cite{BelMul98b}.

As a side remark let us note that in spite of the fact that the conformal
field transformation laws for the dimensionally reduced, from $d = 4 - 2
\varepsilon$ (and $\varepsilon < 0$) to $4$ dimensions, theory differs
from the ones in dimensional regularization by the presence of
$\varepsilon$-scalar contributions, e.g.\footnote{The indices $\mu$,
$\tilde \mu$ and $\bar \mu$ refer to the 4, $d$ and $2 \varepsilon$
dimensional spaces, respectively .} $\tilde\delta^D B^a_\mu = x_{\tilde \nu}
G^a_{\mu\nu} - B^a_{\bar \mu}$ and $\tilde\delta^C_- B^a_\mu
= (2 x_- x_{\tilde\nu} - x^2 n^\star_{\tilde\nu} ) G^a_{\mu\nu}
- 2 x_- B^a_{\bar\mu}$ for the gauge covariant variations of
four-dimensional fields. Nevertheless, the final result for the variation
of the action takes the same form as in Eq.\ (\ref{ConformalVarS}) but
with boson fields being four-dimensional instead.

\section{Superconformal anomalies.}
\label{LargangAnomalies}

In four-dimensional space-time the classical action of the $\cN = 1$
$SU (N_c)$ super-Yang-Mills theory in the Wess-Zumino gauge, i.e.\
\begin{equation}
\label{ClassicLagr}
S_{\rm cl} \equiv \int d^4 x \, \cL_{\rm cl} (x) = \int d^4 x
\left\{ - \frac{1}{4} \left( G^a_{\mu\nu} \right)^2
+ \frac{i}{2} \bar\psi^a \not\!\!\cD^{ab} \psi^b
+ \frac{1}{2} \left( D^a \right)^2 \right\},
\end{equation}
contains the Yang-Mills field strength $G_{\mu\nu}^a = \partial_\mu
B_\nu^a - \partial_\nu B_\mu^a + \g f^{abc} B_\mu^b B_\nu^c$, the Majorana
field $\psi^a$ satisfying the conventional condition $\psi^T C^{(+)} = \bar
\psi$, and an auxiliary field $D^a$. It is invariant under transformation of
the superconformal group which consists besides the conformal group also 
of the translational, $Q$, and conformal, $S$, supertransformations. The
latter two are defined infinitesimally by their action on field operators 
as
\begin{equation}
\label{NonLinSUSY}
\delta^F \psi^a = \frac{i}{2} G^a_{\mu\nu} \sigma_{\mu\nu} \zeta
- i D^a \gamma_5 \zeta,
\qquad
\delta^F B^a_\mu = - i \bar\zeta \gamma_\mu \psi^a,
\qquad
\delta^F D^a = \bar\zeta \not\!\!\cD^{ab}\gamma_5 \psi^b ,
\end{equation}
where $\zeta \equiv \zeta_0 - i\! {\not\! x} \zeta_1$ is the Grassmann
parameter. For $\zeta_1 = 0$ ($F = Q$) we have restricted
supertransformations, while for $\zeta_0 = 0$ ($F = S$) these equations
define the superconformal variations.

The superconformal group is defined by its algebra from which we will be
interested in one particular commutator
\begin{equation}
\label{QKScommutator}
[\cQ, \cK_-]_- = \gamma_- \cS ,
\end{equation}
with $\cQ$ ($\cS$) being super (conformal) generators. Note, however, that
for the short supermultiplet $(B^a_\mu, \psi^a, D^a)$ this commutation
relation is modified for action on fermions. To restore it one has to use
Jackiw's gauge covariant conformal transformation $\widetilde\delta^C_\mu
\equiv \delta^C_\mu + \delta^{\rm gauge}_\mu$ \cite{Jac78}, where the gauge
transformation is defined with field-dependent parameter $\epsilon^a_\mu
\equiv ( 2 x_\mu x_\nu - x^2 g_{\mu\nu} ) B^a_\nu$, instead of the
conventional $\delta^C_\mu$ variation defined above. For the action on a
space spanned by gauge invariant operators this modification is irrelevant.
The commutator (\ref{QKScommutator}), when applied on a Green function with
conformal operator insertion, will provide in the supersymmetric limit
(identifying colour factors) non-trivial relations between the afore
mentioned QCD special conformal anomalies.

To quantize the theory described by the action (\ref{ClassicLagr}) we have
to add a gauge fixing and a ghost term. We do it via the covariant gauge
fixing
\begin{equation}
S_{\rm gf} = \int d^4 x
\left\{
- \frac{1}{2 \xi} (\partial_\mu B^a_\mu)^2
+ \partial_\mu \bar\omega^a \cD^{ab}_\mu \omega^b
\right\}.
\end{equation}
Although it explicitly breaks the supersymmetry on Lagrangian level, it
will not affect gauge invariant quantities since the supersymmetry
variations (\ref{NonLinSUSY}) commute with BRST transformations on the
gauge fixing function\footnote{We write for brevity $\delta^\BRST$ instead
of $\delta^\BRST/\delta\lambda$, i.e.\ after transformation the
infinitesimal Grassmann variable is canceled from the right. We recall
that the BRST transformations are given by the set of equations
$\delta^\BRST B_\mu^a = \cD_\mu^{ab} \omega^b$, $\delta^\BRST \psi^a
= \g f^{abc} \omega^b \psi^c$, $\delta^\BRST \omega^a = \frac{\g}{2} f^{abc}
\omega^b \omega^c$ and $\delta^\BRST \bar\omega^a = \frac{1}{\xi}
\partial_\mu B_\mu^a$.}, i.e.\ $[\delta^F, \delta^\BRST]_- (\partial_\mu
B^a_\mu) = 0$.

Translational and conformal supervariation of the action $S = S_{\rm cl}
+ S_{\rm gf}$ regularized by means of the dimensional regularization,
$4 \to d = 4 - 2 \varepsilon$, leads to
\begin{eqnarray}
\label{deltaQofS}
i \delta^Q S \!\!\!&=&\!\!\!
- \left( \bar\zeta_0 \cO_{3 \psi} \right)
- \cO_Q^{\rm BRST} , \\
i \delta^S S \!\!\!&=&\!\!\!
\frac{d - 4}{2} \left( \bar\zeta_1 \cA \right)
+ \left( \bar\zeta_1 \cO^-_{3 \psi} \right) - \cO_S^{\rm BRST} ,
\end{eqnarray}
where $\cO \equiv \int d^d x \cO (x)$ and the operator insertions read
\begin{eqnarray}
\label{defA}
&&\cA (x) = \sigma_{\mu\nu} G^a_{\mu\nu} \psi^a , \\
\label{O_BRST}
&&\cO_Q^{\rm BRST} (x) =
i \delta^\BRST \delta^Q
\left( \bar\omega^a \partial_{\mu} B^a_{\mu} \right),
\quad
\cO_S^{\rm BRST} (x) = i \delta^\BRST \delta^S
\left( \bar\omega^a \partial_{\mu} B^a_{\mu} \right) , \\
&&\cO_{3 \psi} (x)
= i \frac{\g}{2} f^{abc} \left( \bar\psi^a \gamma_\mu \psi^b \right)
\left( \gamma_\mu \psi^c \right),
\quad
\cO^-_{3 \psi} (x) = \frac{\g}{2} f^{abc}
\left( \bar\psi^a \gamma_\mu \psi^b \right)
\left( \not\!x \gamma_\mu \psi^c \right) ,
\end{eqnarray}
with the operator $\cA$ being the superconformal anomaly \cite{Cur77,Esp85}
in the trace of the supersymmetric current, i.e.\ $\cQ_\rho = \frac{1}{2}
G^a_{\mu\nu} \sigma_{\mu\nu} \gamma_\rho \psi^a$. We used in Eq.\
(\ref{O_BRST}) the identity
\begin{eqnarray*}
\delta^F \delta^{\rm BRST}
\left( \bar\omega^a \partial_\mu B_\mu^a \right)
= 2 \delta^{\rm BRST} \delta^F
\left( \bar\omega^a \partial_\mu B_\mu^a \right)
+ \delta^F {\mit\Omega}_{\bar\omega} .
\end{eqnarray*}
Note that the three-fermion operators $\cO_{3 \psi}$ and $\cO_{3 \psi}^-$
vanish in four dimensions by means of Fierz rearrangement. Moreover,
$\cO^-_{3 \psi}$ can be generated by a special conformal variation of the
operator $\cO_{3 \psi}$, namely, $ \delta_\mu^C \cO_{3 \psi} = i
\gamma_\mu \cO^-_{3 \psi} + \cO (\varepsilon^2)$, where at one-loop
level we can safely neglect the remainder.

Later on we will concentrate on the use of the $\xi = - 3$ gauge in the
derivation of the constraints, which ensures renormalized supersymmetry at
one loop order \cite{MajPogSch80}. This means that the quark and gluon
anomalous dimensions are equal $\gamma_\phi \equiv \gamma_G = \gamma_\psi$.
Furthermore, at one-loop order it was
found that anomaly $\cA$ does not acquire gauge variant counterterms
\cite{Esp85}, provided one uses this particular value of the gauge fixing
parameter. Therefore, we can write to this accuracy $\varepsilon \cA$ in
terms of the renormalized operator
\begin{equation}
\frac{d - 4}{2} \cA
= \frac{\beta_\varepsilon}{\g} [ \cA ] + \cO (\alpha_s^2) ,
\end{equation}
with $d$-dimensional $\beta$-function $\beta_\varepsilon = - \varepsilon \g
+ \beta$.
Finally, we write the superconformal variation of the action to one-loop
accuracy as
\begin{eqnarray}
\label{deltaSofS}
i \delta^S S \!\!\!&=&\!\!\!
 \frac{\beta_\varepsilon}{\g} \bar\zeta_1 [ \cA ]
+ \bar\zeta_1 \cO^-_{3 \psi}  - \cO_S^{\rm BRST}
+ \cO (\alpha_s^2) .
\end{eqnarray}

Let us point out a further consequence of the $\xi = - 3$ gauge, which also
leads to the equality of anomalous dimensions $\gamma_\phi$ and
$\beta$-function, namely $\beta/\g =\gamma_\phi= - \alpha_s/(4 \pi) 3 N_c$.
Consequently, the renormalization of the conformal variation of the action
will be simplified and its integrand in Eq.\ (\ref{ConformalVarS}) reads in
one-loop approximation (see e.g.\ Ref.\ \cite{BelMul98b})
\begin{eqnarray}
&&- \frac{d - 4}{2}
\left( {\cal O}_A (x) + {\cal O}_B (x) + {\mit \Omega}_{\bar \omega} (x)
- {\mit \Omega}_{\bar\psi\psi} (x) - {\mit \Omega}_D (x) \right)
\nonumber\\
&&\qquad\qquad = - \frac{\beta_\varepsilon}{\g}
\left( [\cO_A (x)] - {\mit\Omega}_{\bar\psi\psi} (x) \right)
- \frac{d - 4}{2} ( [\cO_B (x)] + {\mit\Omega}_{\bar\omega} (x)
- {\mit\Omega}_D (x) ) \nonumber\\
&&\qquad\qquad\quad
+ \ 2 \gamma_{\bar\omega} {\mit\Omega}_{\bar\omega} (x)
+ (d - 2) \, \partial_\mu [{\cal O}_{B\mu} (x)] ,
\end{eqnarray}
with ghost anomalous dimension $\gamma_{\bar\omega}$.

\section{Superconformal Ward identities.}
\label{WardIdentitesSC}

In order to derive Ward identities we have to know the change of conformal
operators under the superconformal symmetry. Using the rules in
(\ref{NonLinSUSY}) one finds that the translational supersymmetry
transformation laws are given by (here and everywhere
$\sigma_j = \ft12[1 - (-1)^j]$)
\begin{eqnarray}
\label{WessZuminoMult1row}
&&\delta^Q\, \cS^1_{jl} = \sigma_j\
\bar\zeta_0 \cV_{j - 1 l}, \quad
\delta^Q\, \cS^2_{jl} = \sigma_j\
\bar\zeta_0 \cV_{jl}, \quad
\delta^Q\, \cP^1_{jl} = \sigma_{j + 1}\
\bar\zeta_0 \cU_{j - 1 l}, \quad
\delta^Q\, \cP^2_{jl} = \sigma_{j + 1}\
\bar\zeta_0 \cU_{jl}, \\
\label{WessZuminoMult2row}
&&\delta^Q\, \cV_{j - 1l - 1}
= - \gamma_-\zeta_0
\left\{
\cS^1_{jl} + \cS^2_{j - 1 l}
\right\}
- \gamma_-\gamma_5\zeta_0
\left\{
\cP^1_{jl} + \cP^2_{j - 1 l}
\right\} .
\end{eqnarray}
These equations make our comment about the formation of the conformal
operators into the chiral superfield apparent: Eqs.\
(\ref{WessZuminoMult1row},\ref{WessZuminoMult2row}) are in one-to-one
correspondence with the supersymmetric rules for the Wess-Zumino multiplet
\cite{WesZum70}. Under superconformal variations conformal operators
behave as follows
\begin{eqnarray}
&&\delta^S\, \cS^1_{jl} = - \sigma_j (j + l + 3)\
\bar\zeta_1 \gamma_+ \cV_{j - 1 l - 1}, \qquad
\delta^S\, \cS^2_{jl} = - \sigma_j (l - j)\
\bar\zeta_1 \gamma_+ \cV_{j l - 1}, \nonumber\\
\label{SCvariation12row}
&&\delta^S\, \cP^1_{jl} = - \sigma_{j + 1} (j + l + 3)\
\bar\zeta_1 \gamma_+ \cU_{j - 1 l - 1}, \quad
\delta^S\, \cP^2_{jl} = - \sigma_{j + 1} (l - j)\
\bar\zeta_1 \gamma_+ \cU_{j l - 1}, \\
&&\delta^S\, \cV_{j - 1 l - 1}
= - \gamma_- \gamma_+ \zeta_1
\left\{
(l - j) \cS^1_{j l - 1} + (j + l + 2) \cS^2_{j - 1 l - 1}
\right\} \nonumber\\
&&\qquad\qquad\ \ - \gamma_-\gamma_5 \gamma_+ \zeta_1
\left\{
(l - j) \cP^1_{j l - 1} + (j + l + 2) \cP^2_{j - 1 l - 1}
\right\} .
\label{SCvariation3row}
\end{eqnarray}
Note, that Eqs.\ (\ref{WessZuminoMult1row},\ref{SCvariation12row})
do not require Fierz rearrangement and, therefore, they do not
change their form when the theory is regularized via dimensional
regularization.
Finally, let us recall that the transformation laws of conformal operators
under scaling and special conformal variations are given by
\begin{equation}
\label{ConformalVariation}
\delta^D {^{\mit\Omega}\!\cO_{jl}}
= - \left( l + d({\mit\Omega}) \right) {^{\mit\Omega}\!\cO_{jl}},
\qquad
\delta^C_- {^{\mit\Omega}\!\cO_{jl}}
= i a_{jl}({\mit\Omega}) {^{\mit\Omega}\!\cO_{jl - 1}}.
\end{equation}
Here $d(B)\equiv d(G) = d(Q) = 3$ and $d(F) = \ft72$ as well as $a_{jl}
(B)\equiv a_{jl} (Q)=a_{jl} (G)=a (j,l,1,1)$ and $a_{jl}
(F)=a(j,l,2,1)$ with $a(j,l,\nu_1,\nu_2)=2(j-l)(j+l+\nu_1+\nu_2+1)$,
where $\nu_\phi = d_\phi + s_\phi - 1$ and $s_\phi$ being the spin
of the field $\phi$. Again the scale dimension $d_\phi$ is chosen to
coincide to its canonical value in four dimensions.

Due to difficulties to preserve the supersymmetry of the theory with
quantization and regularization procedures our modest goal will be, 
therefore, a derivation of the constraints for the special conformal 
anomalies of the QCD conformal operators stemming from the commutator 
equation (\ref{QKScommutator}) at one loop level only. We will choose 
the covariant gauge with $\xi = - 3$, which gives us the advantages 
mentioned above.

The dilatation and special conformal Ward identities for a conformal
operator ${^{\mit\Omega}\!\cO}$, which is either bosonic (${\mit\Omega} =
B$) or fermionic (${\mit\Omega} = F$) one, look now very simple (cf.\ Ref.\
\cite{BelMul98b})
\begin{eqnarray}
\label{DWI}
\langle [ {^{\mit\Omega}\!\cO_{jl}} ] \delta^D \cX \rangle
\!\!\!&=&\!\!\! \sum_{k = 0}^{j}
\left\{ \left( l + d({\mit\Omega}) \right) {\bf 1}
+ \mbox{\boldmath$\gamma$} ({\mit\Omega}) \right\}_{jk}
\langle  [ {^{\mit\Omega}\!\cO_{kl}} ] \cX \rangle
+  \frac{\beta}{\g}
\langle i [ {^{\mit\Omega}\!\cO_{jl}}
\left( \cO_A - {\mit \Omega}_{\bar\psi\psi} \right) ] \cX \rangle ,
\\
\label{sCWI}
\langle [ {^{\mit\Omega}\!\cO_{jl}} ] \delta^C_- \cX \rangle
\!\!\!&=&\!\!\! -i \sum_{k = 0}^{j}
\left\{ a_{jl} ({\mit\Omega}) {\bf 1} +
\mbox{\boldmath$\gamma$}^c ({\mit\Omega}) \right\}_{jk}
\langle  [ {^{\mit\Omega}\!\cO_{k l-1}} ] \cX \rangle
+ \frac{\beta}{\g}
\langle i [ {^{\mit\Omega}\!\cO_{jl}}
\left( \cO_A^- - {\mit \Omega}_{\bar\psi\psi}^- \right) ]
\cX \rangle \nonumber\\
&&- 2\langle i [ {^{\mit\Omega}\!\cO_{jl}} \Delta^-_{\rm ext} ]
\cX \rangle ,
\end{eqnarray}
with a two-by-two unit matrix ${\bf 1} \equiv 1_{[2] \times [2]}$.
Here $\mbox{\boldmath$\gamma$}$ and $\mbox{\boldmath$\gamma$}^c$ are
the scale and special conformal anomalies. It is well known, the scale
Ward identity coincides with the Callan-Symanzik equation. Thus,
$\mbox{\boldmath$\gamma$}$ is the conventional anomalous dimension
matrix and the combination $\left( l + d \right) {\bf 1} +
\mbox{\boldmath$\gamma$}$ is the scale dimension matrix of conformal
operators. Obviously $\cO_A (x) - {\mit\Omega}_{\bar\psi\psi} (x) = -
2 \cL_{\rm cl} (x)$ is the classical Lagrangian (\ref{ClassicLagr})
without auxiliary fields. In Eq.\ (\ref{DWI},\ref{sCWI}) we have also
introduced a new conventions for the operator insertions weighed with
different functions, i.e.\ $\cO = \int d^d x\ \cO (x)$, $\cO^- = \int
d^d x\ 2 x_-\ \cO (x)$ (analogously for equation-of-motion operators)
and $\Delta^-_{\rm ext} = \int d^d x\ 2 x_-\ \partial_\mu \cO_{B \mu} (x)$.
The precise definition of the renormalized operator products is given in
the next section.

Let us turn to the renormalized supersymmetric Ward identities. For
definiteness let us consider parity even $\cS$ operators. From
unrenormalized Ward identity (\ref{WardIdentity}), the superconformal
variations of the action (\ref{deltaQofS},\ref{deltaSofS}) and the operators
(\ref{WessZuminoMult1row},\ref{SCvariation12row}) we can immediately derive
the renormalized $Q$ and $S$ supersymmetric Ward identities in one loop
approximation:
\begin{eqnarray}
\label{QsusyWI}
\langle [\cS_{jl}^a] \delta^Q \cX \rangle
\!\!\!&=&\!\!\! - \langle [\delta^Q \cS_{jl}^a] \cX \rangle
+ \langle \left(\bar\zeta_0 \cO_{3\psi} \right) [\cS^a_{jl}]\cX \rangle
+ \langle [\cS_{jl}^a] \cO_Q^{\rm BRST} \cX \rangle , \\
\label{SsusyWI}
\langle [\cS_{jl}^a] \delta^S \cX \rangle
\!\!\!&=&\!\!\! - \langle [\delta^S \cS_{jl}^a] \cX \rangle
+ \sigma_j \bar\zeta_1 \gamma_+ \sum_{k = 0}^{j}
r_{jk}^{a;V [1]}
\langle [\cV_{kl - 1}] \cX \rangle
 - \frac{\beta}{\g}
\langle \left(\bar\zeta_1 \cA \right) [\cS_{jl}^a]  \cX\rangle \nonumber\\
&&\!\!\!
- \langle \left( \bar\zeta_1 \cO_{3\psi}^- \right)
[\cS^a_{jl}] \cX \rangle
+ \langle [\cS_{jl}^a] \cO_S^{\rm BRST} \cX \rangle .
\end{eqnarray}
Here the superconformal anomaly $r^{[1]}_{jk}$ is the residue of the
renormalization constant
\begin{equation}
r_{jk} = r^{[0]}_{jk} + \frac{1}{\varepsilon} r^{[1]}_{jk} + \dots ,
\end{equation}
arisen from the renormalization of the operator product
\begin{equation}
\label{RenA&S}
\left( \bar\zeta_1 [\cA] \right) [\cS^a_{jl}]
= [ \left( \bar\zeta_1 \cA \right) \cS^a_{jl} ]
+ \sigma_j \bar\zeta_1 \gamma_+
\sum_{k = 0}^{j} {^a r^V_{jk}} [\cV_{k,l - 1}]
\end{equation}
and induced by the $\varepsilon$ term of $\beta_\varepsilon$. Note that
since the three-fermion operators vanish in four dimensions, their product
with the bosonic operators will give a finite contribution at one loop
order. Similar equations hold for $\cP_{jl}$ with the replacement of index
$V$ by $A$ and $\sigma_j$ by $\sigma_{j + 1}$.

We have neglected infinite terms in the above Ward identities, since
they have to cancel each other. It is instructive to discuss this issue
in more detail for the supersymmetric Ward identity of the two-vector 
$\mbox{\boldmath$\cS$} = \left( { \cS^1 \atop \cS^2 }\right)$
\begin{equation}
\langle [\mbox{\boldmath$\cS$}_{jl}] \delta^Q \cX \rangle
= - \langle \delta^Q [\mbox{\boldmath$\cS$}_{jl}] \cX \rangle
- \langle i [\mbox{\boldmath$\cS$}_{jl}] (\delta^Q S) \cX \rangle .
\end{equation}
The variation of ``good" component of the fermion field, $\psi_+ \equiv
\ft12 \gamma_- \gamma_+ \psi$, entering in $\cX$ may cause a divergency
on the l.h.s.\, since it contains a composite field strength $G_{\mu\nu}$.
Fortunately, in the light-cone gauge the latter can be expressed in terms
of elementary vector potential $B_\mu$ and, therefore, the l.h.s.\ is
finite by definition. This gauge, together with the use of dimensional
reduction which implies $\delta^Q S = 0$, leads to
\begin{equation}
\label{FinitenessSUSYQ}
\langle \delta^Q [\mbox{\boldmath$\cS$}_{jl}] \cX \rangle = \mbox{finite} .
\end{equation}
Since the renormalization of the composite operators is both gauge and
scheme independent at leading order, Eq.\ (\ref{FinitenessSUSYQ}) holds also
true for our choice of scheme. Furthermore, in Eq.\ (\ref{QsusyWI}) the
renormalization of the product $[\mbox{\boldmath$\cS$}_{jl}] \cO_{3 \psi}$
is finite at leading order because $\cO_{3 \psi} \sim \cO (\varepsilon)$ and
cancels a pole in one-loop diagrams. Consequently, the product
$[\mbox{\boldmath$\cS$}_{jl}] \cO_Q^{\rm BRST} = \mbox{finite}$, or if
divergent it cancels singularities in $[\mbox{\boldmath$\cS$}_{jl}] \delta^Q
\psi$ mentioned above.

Now we discuss the consequences of Eq.\ (\ref{FinitenessSUSYQ}).
The components of the renormalized operators $[\mbox{\boldmath$\cS$}_{jk}]$
are defined in terms of unrenormalized ones, constructed though from the
renormalized fields $\phi$, by
\begin{equation}
\label{RenormSoper}
[\cS^a_{jl}]
= \sum_{b = 1}^2 \sum_{k = 0}^{j} \{ {^{ab}\!Z_\cS} \}_{jk} \, \cS^b_{kl},
\qquad
\cS^a_{kl} = Z_\phi^{- 1} \cS^{a(0)}_{kl},
\end{equation}
where $\mbox{\boldmath$\cS$}^{(0)}_{jl}$ is expressed in terms of the
bare fields $\phi^{(0)} = \sqrt{Z_\phi} \phi$ and coupling $\g^{(0)} =
\mu^\varepsilon \g/\sqrt{Z_\phi}$. The anomalous dimension matrix of the
vector $[\mbox{\boldmath$\cS$}]$ is defined as usual
\begin{equation}
\frac{d}{d \ln \mu} [\mbox{\boldmath$\cS$}]_{jl} =
 \sum_{k = 0}^{j} \mbox{\boldmath$\gamma$}^\cS_{jk}
[\mbox{\boldmath$\cS$}]_{kl},
\quad\mbox{with}\quad
\mbox{\boldmath$\gamma$}^\cS = \mbox{\boldmath$\gamma$}^\cS_Z
+ 2 \gamma_\phi \, {\bf 1}
= \left(
\begin{array}{ll}
{^{11}\gamma} & {^{12}\gamma} \\
{^{21}\gamma} & {^{22}\gamma}
\end{array}
\right) .
\end{equation}
In our scheme the anomalous dimensions
\begin{equation}
\gamma_\phi = \frac{1}{2} \frac{d}{d \ln \mu} \ln Z_\phi
= - \frac{1}{2} \frac{\partial}{\partial \ln \g} Z_\phi^{[1]}
\quad\mbox{and}\quad
\mbox{\boldmath$\gamma$}^\cS_Z
= - \left( \frac{d}{d \ln \mu} \mbox{\boldmath$Z$}_\cS \right)
\mbox{\boldmath$Z$}_\cS^{-1}
= \frac{\partial}{\partial \ln \g} \mbox{\boldmath$Z$}_\cS^{[1]} ,
\end{equation}
are expressed by the residues of the Laurent expansion of the $Z$ factors
$Z = 1 +  Z^{[1]}/\varepsilon + \cO (\varepsilon^{-2})$. The renormalized
fermionic operator is defined by the same equation (\ref{RenormSoper}),
however, with the $2 \times 2$-matrix $\mbox{\boldmath$Z$}_\cS$ being
replaced by the numbers $Z_\cV$.

{\ }From Eq.\ (\ref{FinitenessSUSYQ}) we conclude that
\begin{equation}
\sum_{k = 0}^{j} \sum_{k' = 0}^{k}
\left(
\begin{array}{ll}
\{ {^{11}\!Z}_\cS \}_{jk} & \{ {^{12}\!Z}_\cS \}_{jk} \\
\{ {^{21}\!Z}_\cS \}_{jk} & \{ {^{22}\!Z}_\cS \}_{jk}
\end{array}
\right)
\sigma_k
\left(
\begin{array}{l}
\{ Z_\cV^{-1} \}_{k - 1, k'}  \\
\{ Z_\cV^{-1} \}_{k k'}
\end{array}
\right)
[\cV_{k' l}]
= \mbox{finite} ,
\end{equation}
where we implied that $Z_{jk} = 0$ for $k > j$. Substituting the Laurent
series into this result, the $\frac{1}{\varepsilon}$-poles have to
cancel. This is ensured by the relations
\begin{eqnarray}
\label{Qconstraint}
&&\sum_{k = 0}^{j} \left\{ {^{11}\!Z^{[1]}_\cS} \right\}_{jk}
\, \sigma_k \, [\cV_{k - 1, l}]
+ \sum_{k = 0}^{j} \left\{ {^{12}\!Z^{[1]}_\cS} \right\}_{jk}
\, \sigma_k \, [\cV_{kl}]
= \sigma_j \sum_{k = 0}^{j} \left\{ {Z^{[1]}_\cV} \right\}_{j - 1, k}
[\cV_{kl}] , \\
&&\sum_{k = 0}^{j} \left\{ {^{21}\!Z^{[1]}_\cS} \right\}_{jk}
\, \sigma_k \, [\cV_{k - 1, l}]
+ \sum_{k = 0}^{j} \left\{ {^{22}\!Z^{[1]}_\cS} \right\}_{jk}
\, \sigma_k \, [\cV_{kl}]
= \sigma_j \sum_{k = 0}^{j} \left\{ {Z^{[1]}_\cV} \right\}_{jk}
[\cV_{kl}],
\nonumber
\end{eqnarray}
where $[\cV_{kl}]$ are independent operators. These when combined together
with the analogous results for parity odd operators (replace $\cS \to \cP,\
\cV \to \cU$, and $\sigma_j \to \sigma_{j - 1})$ and ${Z_\cV} = {Z_\cU}$
provide constraints for anomalous dimensions (when differentiated w.r.t.\
$\ln \g$)
\cite{BukFroKurLip85,BelMueSch98}:
\begin{eqnarray}
\label{ConZLO}
&&\left\{ {^{11}\!Z^{[1]}_\cS} \right\}_{2n + 1, 2m + 1}
= \left\{ {^{22}\!Z^{[1]}_\cP} \right\}_{2n, 2m}
= \left\{ {Z_\cV^{[1]}} \right\}_{2n, 2m},
\nonumber\\
&&\left\{ {^{11}\!Z^{[1]}_\cP} \right\}_{2n, 2m}
= \left\{ {^{22}\!Z^{[1]}_\cS} \right\}_{2n - 1, 2m- 1}
= \left\{ {Z_\cV^{[1]}} \right\}_{2n - 1, 2m - 1},
\nonumber\\
&&\left\{ {^{12}\!Z^{[1]}_\cS} \right\}_{2n + 1, 2m + 1}
= \left\{ {^{21}\!Z^{[1]}_\cP} \right\}_{2n, 2m + 2}
= \left\{ {Z^{[1]}_\cV} \right\}_{2n, 2m + 1} ,
\nonumber\\
&&\left\{ {^{21}\!Z^{[1]}_\cS} \right\}_{2n + 1, 2m + 1}
= \left\{ {^{12}\!Z^{[1]}_\cP} \right\}_{2n + 2, 2m}
= \left\{ {Z^{[1]}_\cV} \right\}_{2n + 1, 2m} .
\end{eqnarray}

Now let us turn to the renormalization of the superconformal Ward
identities. An important consequence of these constraints is that the
operators are multiplicatively renormaliziable in the one-loop
approximation, e.g.\ $\left\{{^{12}\!Z^{[1]}_\cS} \right\}_{jj}=
\left\{{^{21}\!Z^{[1]}_\cS} \right\}_{jj}=0$ and $\left\{
{^{11}\!Z^{[1]}_\cS} \right\}_{j+1,j+1} = \left\{ {^{22}\!Z^{[1]}_\cP}
\right\}_{jj} = \left\{ {Z_\cV^{[1]}} \right\}_{jj}$. Thus, in this
approximation the classical transformation laws (\ref{SCvariation12row}) for
superconformal variation remain true also for the renormalized operators.
Consequently, the superconformal variation of the renormalized operator
provide finite Green functions
\begin{equation}
\label{FinitenessSUSYS}
\langle \delta^S [\mbox{\boldmath$\cS$}_{jl}] \cX \rangle =
\langle  [\delta^S \mbox{\boldmath$\cS$}_{jl}] \cX \rangle
= \mbox{finite}.
\end{equation}
Since $\cO_{3\psi}^-$ vanishes in four dimensional space-time,
$\langle \left( \bar\zeta_1\cO_{3\psi}^- \right) [\cS^a_{jl}]\cX
\rangle$ is finite. Thus, also the superconformal Ward identity
(\ref{SsusyWI}) is renormalized up to possible divergencies on the l.h.s.\
that are canceled by the renormalization with BRST-exact operators
on the r.h.s.\ of this Ward identity. Note that in our leading order
approximation the anomalous term proportional to the $\beta$ function is
given by a tree approximation.

\section{Constraint equalities for conformal anomalies.}
\label{ConstraintsSUSY}

Having derived Ward identities we are now able to discuss consequences
of the superconformal algebra. To demonstrate the method, we derive at
first the set of relations (\ref{ConZLO}) between the anomalous dimensions
arising from the commutator of super and scaling variations
\begin{eqnarray}
\label{ComVarQ}
[ \delta^Q ,\delta^D ]_- = \frac{1}{2} \delta^Q,
\end{eqnarray}
which is deduced from the commutator algebra $[\cQ, \cD]_- = \ft{i}2 \cQ$.
Next we deal in the same conceptual manner with the commutator
(\ref{QKScommutator}) of super and special conformal transformations,
written in a symbolical form as
\begin{eqnarray}
\label{ComVarS}
[\delta^Q, \delta^C_-]_- = - i \gamma_- \delta^S.
\end{eqnarray}
This provides us the desired constraints for the
special conformal anomalies of the conformal operators. We mostly
concentrate on the even parity sector and just state the results for
the odd one.

Before we start, let us argue that the three-fermion operators and
BRST-exact operators will not contribute to the constraints. As already
mentioned, the product of the three-fermion operators with composite 
operators provide in one-loop approximation a finite part, which could 
possibly contribute to the constraints. However, its evaluation gives a 
result that depends on the $\gamma_5$ and ${\rm tr} \, {\bf 1}$ 
prescriptions. Let us explain this point in more detail. Calculating the 
contributions of this operator product one deals with a quark loop that 
due to different Wick pairings has three terms: two of them contain a 
string of Dirac matrices while the last one is a trace. Obviously, in 
four dimensions the sum gives zero result, recall that $\cO^{3\psi}$ and 
$\cO^{3\psi}_-$ vanish by means of Fierz rearrangement. However, because 
of $\varepsilon^{-1}$ pole in the loop momentum integration, we have to
keep $\varepsilon$-contributions from the spinor algebra that cancel
this pole. Obviously, the $\varepsilon$ part is ambiguous for axial
channel: the result depends on the handling of $\gamma_5$ in the string
of Dirac matrices as well as in the trace. Next, since one of the
three contributions is given by a trace of Dirac matrices, the result
depends on the prescription for the trace of the unit matrix ${\rm tr} \,
{\bf 1}$. One of the choices made in most calculations is to adopt a
fiction for $d$-dimensional gamma matrices that still ${\rm tr} \, {\bf 1}
= 4$. However, in those computations the trace appears as a single overall
factor and the above choice is permissible. It results into scheme
dependence for finite part of e.g.\ one-loop diagrams. In our case, since
we have an additive trace contribution, we have to continue the Clifford
algebra in $d$ dimensional space as well, this results into the rule
${\rm tr} \, {\bf 1} = 2^{[d/2]}$. This convention produces a term
involving $\ln 2$ reflecting scheme dependence. On the other hand we
certainly know that in leading order the conformal anomalies do not
depend on these ambiguities. So we conclude that the contribution of
the operator products in questions can not affect the constraints.

Now we come to the operator products containing unphysical BRST operators. 
Of course, one expects that these operators do not contribute to the 
physical sector, however, they may be responsible for the cancellation of 
unphysical pieces appearing in the renormalization of products containing 
only gauge invariant operators. On the very end we are interested on
relations for physical quantities and as we already know from the
constraints on anomalous dimensions these operator products have to be die
out in Eq.\ (\ref{ComVarQ}). For the commutator (\ref{ComVarS}), we have the
superconformal anomaly and the only difficulty could be a gauge dependent
term that is cancelled by the operator products in questions. From our
previous experience in Ref.\ \cite{Mue94,BelMul98b}, we expect that such a
contribution is absent and this will be shown by explicit calculation in
section \ref{ExplicitCalcAnom}. So it is justified to neglect the whole
unphysical sector from the very beginning.

\subsection{Commutator constraints for anomalous dimensions.}
\label{ConstaintsQsusy}

First let us demonstrate the derivation of the relations (\ref{ConZLO})
for anomalous dimensions at leading order from the commutator of scale
and supersymmetric variations, given in Eq.\ (\ref{ComVarQ}), applied to
the Green functions of composite operators
\begin{eqnarray}
\label{ConComGreQ}
\langle [\mbox{\boldmath$\cS$}_{jl}]
\left( [\delta^Q, \delta^D] \cX \right) \rangle
= \frac{1}{2} \langle [\mbox{\boldmath$\cS$}_{jl}] \delta^Q \cX \rangle .
\end{eqnarray}
The r.h.s.\ of this equality is obviously given by supersymmetric Ward
identity (\ref{QsusyWI}) times $\ft12$. To calculate the l.h.s.\ of the
commutator we employ the Ward identities (\ref{DWI}) and (\ref{QsusyWI}),
with BRST-exact operators being omitted, and find the following
contributions
\begin{eqnarray}
\label{LHS1comm}
\langle [\mbox{\boldmath$\cS$}_{jl}] \delta^Q \delta^D \cX \rangle
\!\!\!&=&\!\!\! - \frac{\beta}{\g}
\langle i \left[\left(\delta^Q  \mbox{\boldmath$\cS$}_{jl}\right)
\left( \cO_A - {\mit \Omega}_{\bar\psi\psi} \right) \right] \cX \rangle
+ \left( l + \ft72 \right)
\langle \left( \bar\zeta_0 \cO_{3\psi} \right)
[\mbox{\boldmath$\cS$}_{jl}] \cX \rangle
\nonumber\\
&-&\!\!\! \sigma_j \, \bar\zeta_0
\sum_{k = 0}^{j}
\left\{
\left( l + \ft72 \right)
\left(\!\!
\begin{array}{ll}
\delta_{j - 1, k} & 0           \\
0                 & \delta_{jk}
\end{array}
\!\!\right)
+
\left(\!\!
\begin{array}{ll}
\gamma^\cV_{j - 1, k} & 0                \\
0                     & \gamma^\cV_{jk}
\end{array}
\!\!\right)
\right\}
\langle \left(\!\!
\begin{array}{l} {[\cV_{kl}]} \\ {[\cV_{kl}]} \end{array}
\!\!\right) \cX \rangle , \\
\label{LHS2comm}
\langle [\mbox{\boldmath$\cS$}_{jl}] \delta^D \delta^Q \cX \rangle
&=&\!\!\! - \frac{\beta}{\g}
\langle i \delta^Q\left[ \mbox{\boldmath$\cS$}_{jl}
\left( \cO_A - {\mit \Omega}_{\bar\psi\psi} \right) \right] \cX \rangle
+ \left( l + 3 \right)
\langle \left( \bar\zeta_0 \cO_{3\psi} \right)
[\mbox{\boldmath$\cS$}_{jl}] \cX \rangle \nonumber\\
&-&\!\!\! \bar\zeta_0
\sum_{k = 0}^{j} \sigma_k
\left\{
\left( l + 3 \right)
\left(\!\!
\begin{array}{ll}
\delta_{jk} & 0           \\
0                 & \delta_{jk}
\end{array}
\!\!\right)
+
\left(\!\!
\begin{array}{ll}
{^{11}\gamma^\cS_{jk}} & {^{12}\gamma^\cS_{jk}} \\
{^{21}\gamma^\cS_{jk}} & {^{22}\gamma^\cS_{jk}}
\end{array}
\!\!\right)
\right\}
\langle \left(\!\!
\begin{array}{l} {[\cV_{k-1, l}]} \\ {[\cV_{kl}]} \end{array}
\!\!\right) \cX \rangle.
\end{eqnarray}
To derive the r.h.s.\ of Eq.\ (\ref{LHS1comm}) we have used for $\langle
\left( \bar\zeta_0 \cO_{3\psi} \right) [\mbox{\boldmath$\cS$}_{jl}]
\delta^D\cX \rangle $ the scaling Ward identities with $\delta^D \cO_{3\psi}
= - \ft12 \cO_{3\psi}$ and the variation (\ref{ConformalVariation}) of
conformal operators. The variation of the action proportional to $\varepsilon$
is neglected, since this Green function is already finite at one-loop order.
As it was already expected from our previous result (\ref{ConZLO}) and a
discussion about scheme dependence, the three-fermion operator insertion
will not affect the constraints at one loop level.

Now we come to the terms in the above equations proportional to the
$\beta$ function. To derive the r.h.s.\ of (\ref{LHS1comm},\ref{LHS2comm})
we have used an equation (modulo infinite constants which again do not
affect the constraints, since the latter are basically relations between
finite contributions)
\begin{equation}
\label{SLrenormalization}
\langle i [ \mbox{\boldmath$\cS$}_{jl}
\left( \cO_A - {\mit \Omega}_{\bar\psi\psi} \right) ] \cX \rangle
=
\langle i [ \mbox{\boldmath$\cS$}_{jl} ]
\left( [\cO_A] - {\mit \Omega}_{\bar\psi\psi} \right) \cX \rangle
- 2
\langle i [ \mbox{\boldmath$\cS$}_{jl} ] \cX \rangle ,
\end{equation}
to get rid of the variation sign on the field monomial $\delta^Q \cX$.
It results from the study of the differential vertex operator insertions
in the Green function $\langle [\mbox{\boldmath$\cO$}_{jl}] \cX \rangle$
with bosonic conformal operator $\mbox{\boldmath$\cO$} = \left(
{{^Q\!\cO} \atop {^G\!\cO} }\right)$, and one finds \cite{BelMul98b} that
the renormalization constant of the operator product
$[\mbox{\boldmath$\cO$}_{jl}][\cO_A]$,
\begin{eqnarray}
\label{ren-A}
i[{\cal O}_A (x)] [\mbox{\boldmath$\cO$}_{j l}]
= i[{\cal O}_A (x) \mbox{\boldmath$\cO$}_{j l}]
\!\!\!&-&\!\!\!
\delta^{(d)} (x) \sum_{k = 0}^{j}
\left\{ \mbox{\boldmath$Z$}_A \right\}_{jk}
[\mbox{\boldmath$\cO$}_{k l}]
-
\frac{i}{2} \partial_+ \delta^{(d)} (x) \sum_{k = 0}^{j}
\left\{ \mbox{\boldmath$Z$}_A^- \right\}_{jk}
[\mbox{\boldmath$\cO$}_{k l - 1}]
- \dots
\nonumber\\
\!\!\!&-&\!\!\!
\left(
\g \frac{\partial\ln X}{\partial\g}
- 2 \xi \frac{\partial\ln X}{\partial\xi}
\right)
B_\mu^a (x) \frac{\delta}{\delta B_\mu^a (x)}
[\mbox{\boldmath$\cO$}_{j l}] ,
\end{eqnarray}
contains a finite contribution (second term on the r.h.s.)
\begin{equation}
\label{Z_Aconstant}
\mbox{\boldmath$Z$}_A =
\left(
\g \frac{\partial\mbox{\boldmath$Z$}}{\partial\g}
- 2 \xi \frac{\partial\mbox{\boldmath$Z$}}{\partial\xi}
\right) \mbox{\boldmath$Z$}^{-1}
- 2 \mbox{\boldmath$Z$} \mbox{\boldmath$P$}_G \mbox{\boldmath$Z$}^{-1}
- 2
\left(
\g \frac{\partial\ln X}{\partial\g}
- 2 \xi \frac{\partial\ln X}{\partial\xi}
\right)
\mbox{\boldmath$Z$} \mbox{\boldmath$P$}_G \mbox{\boldmath$Z$}^{-1} .
\end{equation}
The constant $X$ is related to the charge and gluon wave function
renormalization constants by the relation $X = Z_\g \sqrt{Z_G}$.
For the $[\mbox{\boldmath$\cS$}_{jl}] {\mit\Omega}_{\bar\psi\psi}$ we
have to use the identity
\begin{equation}
\label{ren-EOM}
i  [ {\cal O}_{jl}] {\mit \Omega}_\phi(x) =  i  [ {\cal O}_{jl}
{\mit \Omega}_\phi(x)]
-\phi(x) \frac{\delta}{\delta \phi(x)} [ {\cal O}_{jl}].
\end{equation}
where it is obvious that
\begin{equation}
\label{ren-EOM-1}
\langle [ {\cal O}_{jl} {\mit\Omega}_\phi(x)] \cX \rangle
= i \langle
[ {\cal O}_{jl}] \phi (x) \frac{\delta}{\delta \phi(x)} \cX
\rangle.
\end{equation}
So that in Eq.\ (\ref{SLrenormalization}) the finite piece appears as a
consequence of two contributions: ${^G\!\cO}$ part from the
$[\mbox{\boldmath$\cS$}_{jl}][\cO_A]$ product due to finite part in
Eqs.\ (\ref{ren-A},\ref{Z_Aconstant}) and ${^Q\!\cO}$ part from the
$[\mbox{\boldmath$\cS$}_{jl}]{\mit\Omega}_{\bar\psi\psi}$ by means of
Eq.\ (\ref{ren-EOM}). Similarly, we have for the fermion operators a
finite contribution
\begin{equation}
\label{VO_Afinite}
i [\cV_{jl}][\cO_A] = i [\cV_{jl} \cO_A] + [\cV_{jl}]
+ \cO (\varepsilon^{- r}) ,
\end{equation}
for the trace anomaly, and
\begin{equation}
\label{VEOMfinite}
i [\cV_{jl}] {\mit\Omega}_{\bar\psi \psi}
= i [\cV_{jl} {\mit\Omega}_{\bar\psi \psi}] - [\cV_{jl}]
+ \cO (\varepsilon^{- r}) ,
\end{equation}
for the equation-of-motion insertion. In both cases one should note the
factors of one in front of the second term on the r.h.s.\, not 2 as for
the quark and gluon operators. Although ${\mit\Omega}_{\psi} =
{\mit\Omega}_{\bar\psi}$ for Majorana fermions (recall $\bar\psi^a
\gamma_\mu \psi^a = 0$ due to the Majorana flip properties), $\psi$
and $\bar\psi$ are treated as independent variables in the functional
integral. Finally, we have
\begin{equation}
\label{ScalePhysSector}
\langle i \delta^Q [ \mbox{\boldmath$\cS$}_{jl}
\left( \cO_A - {\mit \Omega}_{\bar\psi\psi} \right) ] \cX \rangle
=
\langle i [ \left( \delta^Q \mbox{\boldmath$\cS$}_{jl} \right)
\left( \cO_A - {\mit \Omega}_{\bar\psi\psi} \right) ] \cX \rangle ,
\end{equation}
which is almost a trivial result. To derive it one uses
(\ref{SLrenormalization}) and observe that the variation of the
last term $\langle [\mbox{\boldmath$\cS$}_{jl}] \cX \rangle$ in it cancels
with the second terms in Eqs.\ (\ref{VO_Afinite},\ref{VEOMfinite}) so
that we are left with the r.h.s.\ of (\ref{ScalePhysSector}). All other
terms in the commutator of Ward identities are relatively straightforward
to handle. Obviously, with the equality (\ref{ScalePhysSector}) the
contributions proportional to the $\beta$ function in Eqs.\ (\ref{LHS1comm})
and (\ref{LHS2comm}) cancel each other in the commutator relation.

Subtracting Eq.\ (\ref{LHS1comm}) from (\ref{LHS2comm}) and comparing the
difference with the supersymmetric Ward identity (\ref{QsusyWI}) we get,
after extraction of independent combinations and identifying both parity
sectors for fermionic operators, the known supersymmetric relations
\cite{BelMueSch98} (see also \cite{BukFroKurLip85}):
\begin{eqnarray}
\label{SUSYrelations}
{^{11}\!\gamma}^{\cal S}_{2n + 1, 2m + 1}
\!\!\!&=&\!\!\! {^{22}\!\gamma}^{\cal P}_{2n, 2m}
= \gamma^\cV_{2n, 2m} , \quad m \leq n , \nonumber\\
{^{12}\!\gamma}^{\cal S}_{2n + 1, 2m + 1}
\!\!\!&=&\!\!\! {^{21}\!\gamma}^{\cal P}_{2n, 2m + 2}
= \gamma^\cV_{2n, 2m + 1} , \quad  m \leq n - 1 , \nonumber\\
{^{21}\!\gamma}^{\cal S}_{2n + 1, 2m + 1}
\!\!\!&=&\!\!\! {^{12}\!\gamma}^{\cal P}_{2n + 2, 2m}
= \gamma^\cV_{2n + 1, 2m} , \quad m \leq n \ , \\
{^{22}\!\gamma}^{\cal S}_{2n + 1, 2m + 1}
\!\!\!&=&\!\!\! {^{11}\!\gamma}^{\cal P}_{2n + 2, 2m + 2}
= \gamma^\cV_{2n + 1, 2m + 1} , \quad m \leq n \ , \nonumber\\
{^{12}\!\gamma}^{\cal S}_{2n + 1, 2n + 1}
\!\!\!&=&\!\!\! 0 \ , \quad
{^{12}\!\gamma}^{\cal P}_{2n, 2n} = 0 .
\nonumber
\end{eqnarray}
Obviously, these are the same relations as given in Eq.\ (\ref{ConZLO}).
If one relies on a supersymmetry preserving scheme, these equations can
be derived in any order of perturbation theory and they have been checked
at two-loop order\footnote{Here we evaluated the rotation matrices from
the conventional $\overline{\rm MS}$, in which all next-to-leading
anomalous dimensions are available, to dimensional reduction scheme. As
mentioned in the introduction this procedure does not completely fix the
non-diagonal part of the rotation matrices.} \cite{BelMueSch98}.

\subsection{Commutator constraints for special conformal anomalies.}
\label{ConstaintsSsusy}

The constraints for the special conformal anomalies of the conformal
operators result from the commutator (\ref{ComVarS}) of supersymmetric and
special conformal variations applied to the Green function:
\begin{eqnarray}
\label{ConComGreS}
\langle [\mbox{\boldmath$\cS$}_{jl}]
\left([\delta^Q, \delta^D] \cX\right) \rangle
= - i \gamma_- \langle [\mbox{\boldmath$\cS$}_{jl}] \delta^S \cX \rangle.
\end{eqnarray}
The derivation runs along the same line as above up to the appearance of the
superconformal anomaly on the r.h.s., given by the Ward identity
(\ref{SsusyWI}), and the absence of finite contributions to the
renormalization of the product $[\mbox{\boldmath$\cO$}_{jl}][\cO_A^-]$.
Again omitting the BRST-exact operator insertions, the commutator of
the l.h.s.\ is given by the two equations
\begin{eqnarray}
\label{LHS1commCon}
\langle [\mbox{\boldmath$\cS$}_{jl}] \delta^Q \delta^C_- \cX \rangle
\!\!\!&=&\!\!\! - \frac{\beta}{\g}
\langle i \left[ \left(\delta^Q \mbox{\boldmath$\cS$}_{jl}\right)
\left( \cO_A^- - {\mit \Omega}_{\bar\psi\psi}^- \right) \right] \cX \rangle
+ \langle [ \left( \bar\zeta_0 \cO_{3\psi}\right)
\mbox{\boldmath$\cS$}_{jl}] \delta_-^C \cX \rangle
\nonumber\\
&-& \!\!\! i \sigma_j \, \bar\zeta_0
\sum_{k = 0}^{j}
\left\{ a_{jl}(F)
\left(\!\!
\begin{array}{ll}
\delta_{j - 1, k} & 0           \\
0                 & \delta_{jk}
\end{array}
\!\!\right)
+
\left(\!\!
\begin{array}{ll}
\gamma^{c,\cV}_{j - 1, k} & 0                \\
0                     & \gamma^{c,\cV}_{jk}
\end{array}
\!\!\right)
\right\}
\langle \left(\!\!
\begin{array}{l} {[\cV_{k,l-1}]} \\ {[\cV_{k,l-1}]} \end{array}
\!\!\right) \cX \rangle , \\
\label{LHS2commCon}
\langle [\mbox{\boldmath$\cS$}_{jl}] \delta^C_- \delta^Q \cX \rangle
\!\!\! &=&\!\!\! - \frac{\beta}{\g}
\langle i \delta^Q \left[\mbox{\boldmath$\cS$}_{jl}
\left( \cO_A^- - {\mit \Omega}_{\bar\psi\psi}^- \right) \right] \cX \rangle
- i a_{jl}(B) \langle [ \left( \bar\zeta_0 \cO_{3\psi} \right)
\mbox{\boldmath$\cS$}_{j, l - 1}] \cX \rangle
\nonumber\\
&-&\!\!\! i\bar\zeta_0
\sum_{k = 0}^{j} \sigma_k
\left\{
a_{jl}(B)
\left(\!\!
\begin{array}{ll}
\delta_{jk} & 0           \\
0                 & \delta_{jk}
\end{array}
\!\!\right)
+
\left(\!\!
\begin{array}{ll}
{^{11}\gamma^{c,\cS}_{jk}} & {^{12}\gamma^{c,\cS}_{jk}} \\
{^{21}\gamma^{c,\cS}_{jk}} & {^{22}\gamma^{c,\cS}_{jk}}
\end{array}
\!\!\right)
\right\}
\langle \left(\!\!
\begin{array}{l} {[\cV_{k-1, l-1}]} \\ {[\cV_{k,l-1}]} \end{array}
\!\!\right) \cX \rangle.
\end{eqnarray}
The conformal variation of the Green function with three-fermion
operator appearing in Eq.\ (\ref{LHS1commCon}) can be calculated at
tree level:
\begin{eqnarray*}
\langle [ \left( \bar\zeta_0 \cO_{3\psi} \right)
\mbox{\boldmath$\cS$}_{jl}]\delta_-^C \cX \rangle
= - i a_{jl}(B) \langle [\left( \bar\zeta_0 \cO_{3\psi} \right)
\mbox{\boldmath$\cS$}_{j, l - 1}] \cX \rangle
- \langle [ \left( \bar\zeta_0 \delta_-^C\cO_{3\psi} \right)
\mbox{\boldmath$\cS$}_{jl}] \cX \rangle .
\nonumber
\end{eqnarray*}
Here we again neglected the BRST-exact operator that arises from the
conformal variation of the action. Taking into account the Ward
identity (\ref{SsusyWI}) and the relation $\delta_-^C \cO_{3 \psi} = i
\gamma_- \cO^-_{3 \psi}$, we observe again that the three-fermion operator
contributions cancel each other in the commutator constraint.

Let us now consider the operator product proportional to the $\beta$
function, which is defined by \cite{BelMul98b}
\begin{equation}
\label{OopCren}
\langle i \left[ \mbox{\boldmath$\cO$}_{jl}
\left( \cO_A^- - {\mit \Omega}_{\bar\psi\psi}^- \right) \right] \cX \rangle
=
\langle i [\mbox{\boldmath$\cO$}_{jl}]
[\left( \cO_A^- - {\mit \Omega}_{\bar\psi\psi}^- \right)] \cX \rangle
- \langle
\left( \int d^d x 2 x_-
\left(
\psi \frac{\delta}{\delta\psi} + \bar\psi \frac{\delta}{\delta\bar\psi}
\right)
[\mbox{\boldmath$\cO$}_{jl}] \right) \cX
\rangle .
\end{equation}
The supersymmetric variation of this product gives
\begin{eqnarray}
\label{SUSYvarofConf}
&&\langle i \delta^Q \left[ \mbox{\boldmath$\cS$}_{jl}
\left( \cO_A^- - {\mit \Omega}_{\bar\psi\psi}^- \right) \right] \cX \rangle
=
\langle
i [\left( \delta^Q \mbox{\boldmath$\cS$}_{jl} \right)
\left( \cO_A^- - {\mit \Omega}_{\bar\psi\psi}^- \right)] \cX \rangle
+ i \langle [\left( \bar\zeta_0 \gamma_- \cA \right)
\mbox{\boldmath$\cS$}_{jl}] \cX \rangle
\nonumber\\
&&\qquad+
\langle
\left( \int d^d x 2 x_- \psi \frac{\delta}{\delta\psi} \
\delta^Q [\mbox{\boldmath$\cS$}_{jl}] \right) \cX
\rangle
- \langle
\delta^Q \left( \int d^d x 2 x_-
\left(
\psi \frac{\delta}{\delta\psi} + \bar\psi \frac{\delta}{\delta\bar\psi}
\right)
[\mbox{\boldmath$\cS$}_{jl}] \right) \cX
\rangle .
\end{eqnarray}
Here we have used $\delta^Q \left( \cO_A^- - {\mit\Omega}_{\bar\psi\psi}^-
\right) = \bar\zeta_0 \gamma_- \cA + \dots$, with ellipsis standing for the
$\cO_{3\psi}$ operator which again is irrelevant in one-loop approximation,
since the whole contribution (\ref{SUSYvarofConf}) is multiplied by the
$\beta$-function and thus starts from $\alpha_s$. The first line of this
equation ensures the cancellation with the same terms appearing in Eqs.\
(\ref{SsusyWI},\ref{LHS1commCon}), however, the remaining equation of motion
operators in the second line will contribute to the constraints for the
special conformal anomaly. We can absorb these additional pieces by a
redefinition of the special conformal anomaly matrix. The action of the
equation-of-motion operators is
\begin{eqnarray}
\label{FermiBmatrix}
&&\int d^d x \ 2 x_-
\left(
\psi \frac{\delta}{\delta\psi} + \bar\psi \frac{\delta}{\delta\bar\psi}
\right){^Q\! \cO}_{jl}
= 2i \sum_{k = 0}^j b_{jk}(B)\, {^Q\! \cO}_{k, l - 1}, \nonumber\\
&&\int d^d x \ 2 x_- \psi \frac{\delta}{\delta\psi} \cV_{jl}
= 2i \sum_{k = 0}^j b_{jk}(F)\, \cV_{k, l - 1},
\end{eqnarray}
where $b_{jk}(B)$ matrix can be found in \cite{Mue94,BelMul98b} and the
fermionic one is evaluated in Appendix \ref{matrixbF}. Now we shift the
$QQ$-entry by ${^{QQ}\!\gamma^{c}_{jk}} \to {^{QQ}\!{\mit\Gamma}^{c}_{jk}}
\equiv {^{QQ}\!\gamma^{c}_{jk}} + 2 \frac{\beta}{\g} b_{jk} (B)$ and set for
the remaining channels ${^{AB}\!{\mit\Gamma}^{c}_{jk}} \equiv
{^{AB}\!\gamma^{c}_{jk}}$. This redefinition treats the $\beta$ term
equivalently for quarks and gluons and instead of Eq.\
(\ref{QGspecialConfAnom}) we have at leading order
\begin{equation}
a_{jk}^{- 1} (B) \mbox{\boldmath${\mit\Gamma}$}^{c(0)}_{jk} (B)
\equiv
- \mbox{\boldmath$d$}_{jk}
\left( \mbox{\boldmath$\gamma$}^{{\rm D}(0)}_k
- \beta_0 {\bf 1} \right)
+ \mbox{\boldmath$g$}_{jk} .
\end{equation}
The new special conformal anomaly matrix
\begin{equation}
\mbox{\boldmath${\mit\Gamma}$}^{c}
=
\left(
\begin{array}{ll}
{^{11}\!{\mit\Gamma}^{c}} & {^{12}\!{\mit\Gamma}^{c}} \\
{^{21}\!{\mit\Gamma}^{c}} & {^{22}\!{\mit\Gamma}^{c}}
\end{array}
\right)
\end{equation}
for the $\mbox{\boldmath$\cS$}_{jl}$ operators can be found from
the conventional quark and gluon ones by the
transformation
\begin{equation}
\label{abTOQGrotation}
\left(
\begin{array}{r}
\frac{1}{k}\,     {^{11}\!{\mit\Gamma}^{c}_{jk}} \\
\frac{1}{k + 1}\, {^{12}\!{\mit\Gamma}^{c}_{jk}} \\
\frac{1}{k}\,     {^{21}\!{\mit\Gamma}^{c}_{jk}} \\
\frac{1}{k + 1}\, {^{22}\!{\mit\Gamma}^{c}_{jk}}
\end{array}
\right)
= \frac{1}{2k + 3}
\left(
\begin{array}{cccc}
1                     & \frac{k + 3}{6}
& \frac{6}{j}         & \frac{k + 3}{j} \\
-1                    & \frac{k}{6}
& - \frac{6}{j}       & \frac{k}{j} \\
- \frac{j + 3}{j + 1} & - \frac{(k + 3)(j + 3)}{6(j + 1)}
& \frac{6}{j + 1}     & \frac{k + 3}{j + 1} \\
\frac{j + 3}{j + 1}   & - \frac{k(j + 3)}{6(j + 1)}
& - \frac{6}{j + 1}   & \frac{k}{j + 1}
\end{array}
\right)
\left(
\begin{array}{r}
{^{QQ}\!{\mit\Gamma}^{c}_{jk}} \\
{^{QG}\!{\mit\Gamma}^{c}_{jk}} \\
{^{GQ}\!{\mit\Gamma}^{c}_{jk}} \\
{^{GG}\!{\mit\Gamma}^{c}_{jk}}
\end{array}
\right) .
\end{equation}
An analogous convention is introduced for the special conformal anomaly
matrix of the fermionic operators\footnote{
Here the fermionic conformal anomaly $\gamma^{c} (F)$ has the same
structure as the quark one, namely, $\gamma^{c} (F) = - b (F) \gamma (F)
+ w (F)$, with anomalous dimension $\gamma (F)$ of the quark-gluon operator
${\cal V}$ (${\cal U}$) and a part $w (F)$ deduced from the renormalization
of ${\cal V}$ (${\cal U}$) with the trace of the energy-momentum tensor. }
\begin{equation}
{\mit\Gamma}^{c}_{jk} (F)
\equiv
\gamma^{c}_{jk} (F) + 2 \frac{\beta}{\g} b_{jk} (F) .
\end{equation}

If one compares now both sides of Eq.\ (\ref{ConComGreS}) in terms of
these new conventions we find that besides the special conformal anomaly
${\mit\Gamma}^{c}$ only the superconformal anomaly,
\begin{equation}
\label{DeltaAnomaly}
{^a\!{\mit\Delta}^i_{jk}} \equiv - 2\, {^a r^{i [1]}_{jk}} ,
\end{equation}
arising from the renormalization of $[\cA] \mbox{\boldmath$\cS$}_{jl}$
in Eq.\ (\ref{RenA&S}), contributes to the desired constraints:
\begin{eqnarray}
\label{SCconstraint1}
&&\sum_{k = 0}^{j} {^{11}\!{\mit\Gamma}^{c, V}_{jk}} (B) \sigma_k
[\cV_{k - 1, l - 1}]
+ \sum_{k = 0}^{j} {^{12}\!{\mit\Gamma}^{c, V}_{jk}} (B) \sigma_k
[\cV_{k l - 1}] \nonumber\\
&&\qquad\qquad\qquad\qquad\qquad\qquad\qquad\qquad\qquad
- \sigma_j \sum_{k = 0}^{j}
\left\{
{\mit\Gamma}^{c}_{j - 1, k} (F) -  {^1\!\!{\mit\Delta}^V_{jk}}
\right\}
[\cV_{k l - 1}] = 0 , \\
\label{SCconstraint2}
&&\sum_{k = 0}^{j} {^{22}\!{\mit\Gamma}^{c, V}_{jk}} (B) \sigma_k
[\cV_{k l - 1}]
+ \sum_{k = 0}^{j} {^{21}\!{\mit\Gamma}^{c, V}_{jk}} (B) \sigma_k
[\cV_{k - 1, l - 1}] \nonumber\\
&&\qquad\qquad\qquad\qquad\qquad\qquad\qquad\qquad\qquad
- \sigma_j \sum_{k = 0}^{j}
\left\{
{\mit\Gamma}^{c}_{jk} (F) - {^2\!\!{\mit\Delta}^V_{jk}}
\right\}
[\cV_{k l - 1}] = 0 ,
\end{eqnarray}
and for $\cP_{jl}$ they read
\begin{eqnarray}
\label{SCconstraint3}
&&\sum_{k = 0}^{j} {^{11}\!{\mit\Gamma}^{c, A}_{jk}} (B) \sigma_{k + 1}
[\cU_{k - 1, l - 1}]
+ \sum_{k = 0}^{j} {^{12}\!{\mit\Gamma}^{c, A}_{jk}} (B) \sigma_{k + 1}
[\cU_{k l - 1}] \nonumber\\
&&\qquad\qquad\qquad\qquad\qquad\qquad\qquad\qquad\qquad
- \sigma_{j + 1} \sum_{k = 0}^{j}
\left\{
{\mit\Gamma}^{c}_{j - 1, k} (F) - {^1\!\!{\mit\Delta}^A_{jk}}
\right\}
[\cU_{k l - 1}] = 0 , \\
\label{SCconstraint4}
&&\sum_{k = 0}^{j} {^{22}\!{\mit\Gamma}^{c, A}_{jk}} (B) \sigma_{k + 1}
[\cU_{k l - 1}]
+ \sum_{k = 0}^{j} {^{21}\!{\mit\Gamma}^{c, A}_{jk}} (B) \sigma_{k + 1}
[\cU_{k - 1, l - 1}]\nonumber\\
&&\qquad\qquad\qquad\qquad\qquad\qquad\qquad\qquad\qquad
- \sigma_{j + 1} \sum_{k = 0}^{j}
\left\{
{\mit\Gamma}^{c}_{jk} (F) - {^2\!\!{\mit\Delta}^A_{jk}}
\right\}
[\cU_{k l - 1}] = 0,
\end{eqnarray}
where we have implied ${\mit\Gamma}^{c}_{jk} = 0$ for $k > j$.
Extracting the independent components from Eqs.\
(\ref{SCconstraint1}-\ref{SCconstraint4}) we finally obtain four equalities
for the non-diagonal elements
\begin{eqnarray}
\label{ConstraintFirst}
&&\!\!\!\!\!\!\!\!\!\!{^{11}\!{\mit\Gamma}^{c, V}_{2n + 1, 2m + 1}} (B)
+ {^1\!\!{\mit\Delta}^V_{2n + 1, 2m}}
=
{^{22}\!{\mit\Gamma}^{c, A}_{2n, 2m}} (B)
+ {^2\!\!{\mit\Delta}^A_{2n, 2m}}
=
{\mit\Gamma}^{c}_{2n, 2m} (F),
\\
&&\!\!\!\!\!\!\!\!\!\!{^{22}\!{\mit\Gamma}^{c, V}_{2n + 1, 2m + 1}} (B)
+ {^2\!\!{\mit\Delta}^V_{2n + 1, 2m + 1}}
=
{^{11}\!{\mit\Gamma}^{c, A}_{2n + 2, 2m + 2}} (B)
+ {^1\!\!{\mit\Delta}^A_{2n + 2, 2m + 1}}
=
{\mit\Gamma}^{c}_{2n + 1, 2m + 1} (F),
\\
&&\!\!\!\!\!\!\!\!\!\!{^{12}\!{\mit\Gamma}^{c, V}_{2n + 1, 2m + 1}} (B)
+ {^1\!\!{\mit\Delta}^V_{2n + 1, 2m + 1}}
=
{^{21}\!{\mit\Gamma}^{c, A}_{2n, 2m + 2}} (B)
+ {^2\!\!{\mit\Delta}^A_{2n, 2m + 1}}
=
{\mit\Gamma}^{c}_{2n, 2m + 1} (F),
\\
&&\!\!\!\!\!\!\!\!\!\!{^{21}\!{\mit\Gamma}^{c, V}_{2n + 1, 2m + 1}} (B)
+ {^2\!\!{\mit\Delta}^V_{2n + 1, 2m}}
=
{^{12}\!{\mit\Gamma}^{c, A}_{2n + 2, 2m}} (B)
+ {^1\!\!{\mit\Delta}^A_{2n + 2, 2m}}
=
{\mit\Gamma}^{c}_{2n + 1, 2m} (F),
\label{ConstraintLast}
\end{eqnarray}
with $n > m$, and six equations for the diagonal elements, which are of no
relevance in prediction (\ref{PreAnoDim}) for the anomalous dimension
matrix. Eqs.\ (\ref{ConstraintFirst}-\ref{ConstraintLast}) are the main
results of this section. The rest of the paper is devoted to a consistency
check of our results for the special conformal anomalies by evaluating the
${\mit\Delta}$-anomalies at one-loop order.

\section{Evaluation of superconformal anomaly.}
\label{ExplicitCalcAnom}


\begin{figure}[t]
\begin{center}
\vspace{0.6cm}
\hspace{1.2cm}
\mbox{
\begin{picture}(0,220)(270,0)
\put(0,-30)                    {
\epsffile{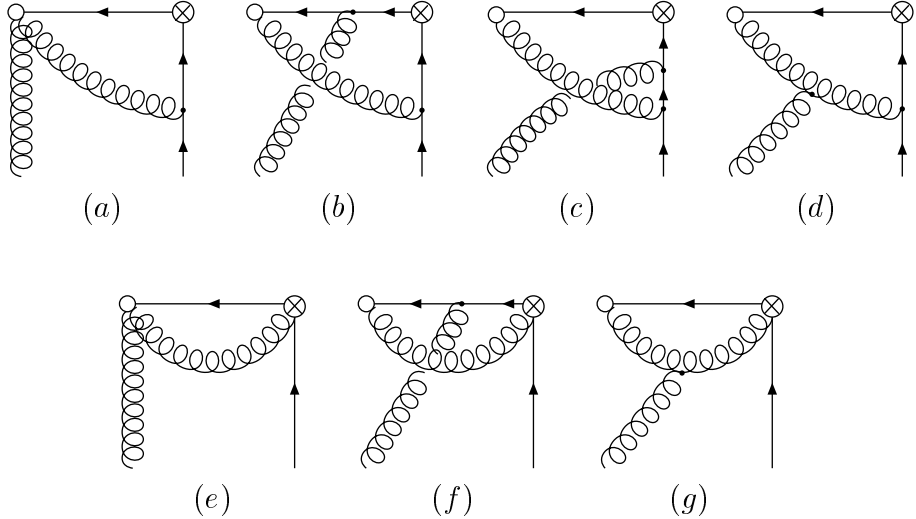}
                               }
\end{picture}
}
\end{center}
\vspace{-2cm}
\caption{\label{RAquarks} One-loop diagrams which give rise to divergences
in the product of the renormalized operator insertions
$i [ \cA ][ {^Q\!\cO_{jl}} ]$.}
\end{figure}


\begin{figure}[t]
\begin{center}
\vspace{0.6cm}
\hspace{1.2cm}
\mbox{
\begin{picture}(0,220)(270,0)
\put(0,-30)                    {
\epsffile{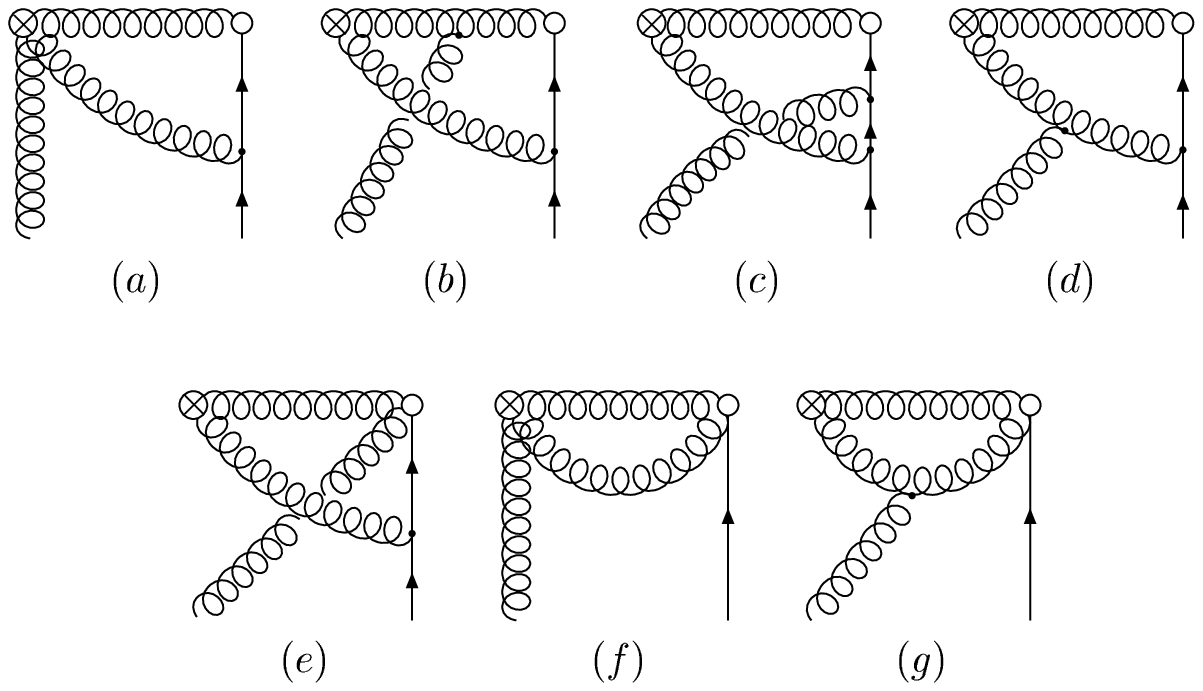}
                               }
\end{picture}
}
\end{center}
\vspace{-2cm}
\caption{\label{RAgluons} Same as in Fig.\ \protect\ref{RAquarks} but
for the operator product $i [ \cA ][ {^G\!\cO_{jl}} ]$.}
\end{figure}

To simplify the calculation of the renormalization constants 
${\mit\Delta}$, it is advantageous to use the light-cone position 
formalism and to this end we introduce the non-local light-ray operators
\begin{eqnarray}
{^Q\!\cO^i} (\kappa_1, \kappa_2)
\!\!\!&=&\!\!\! \frac{1}{2} \bar\psi^a_+ (\kappa_2 n)
{\mit\Gamma}^i {\mit\Phi}^{ab} [\kappa_2, \kappa_1]
\psi^b_+ (\kappa_1 n) , \nonumber\\
{^G\!\cO^i} (\kappa_1, \kappa_2)
\!\!\!&=&\!\!\! G^a_{+\mu} (\kappa_2 n) \cT^i_{\mu\nu}
{\mit\Phi}^{ab} [\kappa_2, \kappa_1]
G^b_{\nu +} (\kappa_1 n) , \nonumber\\
{^F\!\cO^i} (\kappa_1, \kappa_2)
\!\!\!&=&\!\!\! G^a_{+ \mu} (\kappa_2 n) \cF^i_\mu
{\mit\Phi}^{ab} [\kappa_2, \kappa_1]
\psi^b_+ (\kappa_1 n) .
\end{eqnarray}
The calculations of the mixing kernels for the product $[\cA] [\cO]$
with fermionic operators are straightforward and go along the line of Ref.\
\cite{BraHorRob84} (see also appendix B of Ref.\ \cite{BelMul98b} for a
recent review of this formalism in the case when the non-forwardness is
essential). The diagrams are represented in Figs.\ \ref{RAquarks}
and \ref{RAgluons} for quark and gluon operators, respectively.
Since these renormalization constants must be gauge-independent,
we have performed the computation with a general $\xi$ and indeed
observed its cancelation in the final result. This shows that
the BRST-exact operators can not contribute to the constraints. The
generic form of the result is
\begin{eqnarray}
[{^{\mit\Omega}\!\cO^i}] (\kappa_1, \kappa_2)
\left( \bar\zeta_1 [\cA] \right)
&=& \frac{1}{\varepsilon} \frac{\alpha_s}{2 \pi} N_c
\int_{0}^{1} dx \int_{0}^{1} dy \theta (1 - x - y)
{^{\mit\Omega} \cR^i} (x, y)
\nonumber\\
&&\qquad\qquad\times \left( \bar\zeta_1 \gamma_+ {^F\!\cO^i} \right)
( \bar x \kappa_1 + x \kappa_2, y \kappa_1 + \bar y \kappa_2 )
\mp (\kappa_1 \leftrightarrow \kappa_2) ,
\label{OAnonLocEE}
\end{eqnarray}
where the $- (+)$-sign in the second term corresponds to the parity even
quarks and parity odd gluons (parity even gluons and parity odd quarks)
and we use throughout the convention $\bar x \equiv 1 - x$. Skipping
details (see Appendix \ref{Diagram-by-diagram}) we give the result
\begin{eqnarray}
\label{RAVkernelsQUARK}
\!\!\!\!\!- k_{2+} {^Q \cR^V} (x, y) \!\!\!&=&\!\!\!
- \left[ \frac{1}{y} \right]_+ \delta (x),
\qquad\qquad\qquad\qquad\qquad\qquad
{^Q \cR^A} (x, y) = {^Q \cR^V} (x, y) , \\
\label{RAVkernelsGLUON}
\!\!\!\!\!{^G \cR^V} (x, y) \!\!\!&=&\!\!\!
- \left[ \frac{1}{x} \right]_+ \delta (y)
- 2 + \delta (y) + (1 - y) \delta (x), \quad
{^G \cR^A} (x, y) = {^G \cR^V} (x, y) + 4 y ,
\end{eqnarray}
with the $+$-prescription defined conventionally by
\begin{eqnarray*}
\left[ \frac{1}{x} \right]_+ = \frac{1}{x}
- \delta (x) \int_{0}^{1} \frac{dy}{y} .
\end{eqnarray*}
We have kept in the quark sector the gluon momentum $k_{2+}$ unintegrated,
which stems from the exponential in the Feynman rules for the operator
insertion ${\rm Vertex}\times \exp \left( - i \kappa_1 k_{1+} - i \kappa_2
k_{2+} \right)$. We merely substitute it by the corresponding parton
momentum fraction when passing to exclusive type kernels (see Appendix
\ref{Moments}). Note that the result for the polarized quark kernel is
the same as for the parity even case. This could be anticipated, since
for divergent parts we can use anticommutativity of $\gamma_5$.
However, there is a sign change in the contribution of the diagram in Fig.\
\ref{RAquarks} (c) but it gives the contribution $\pm x \delta (1 - x - y)$
with $+(-)$ sign for even (odd) parity and together with $\bar x \delta
(1 - x - y)$ from the Fig.\ \ref{RAquarks} (d) gives a vanishing result
when the symmetry property of the corresponding quark string operators is
used. For gluons the difference between the parity even and odd sectors
comes entirely from the contributions of the graphs in Figs.\ \ref{RAgluons}
(b), (c) and (e). The contributions from (d) and (g) annihilate each
other, while remaining diagrams give identical\footnote{Discarding the
fact that contributions with $\kappa_1 \leftrightarrow \kappa_2$ enter
with opposite signs.} results for $V$ and $A$ channels.

Evaluation of the conformal moments according to the method spelled
in Appendix \ref{Moments} gives for gluons
\begin{equation}
\label{GluonR}
\left( \frac{\alpha_s}{2 \pi} N_c \right)^{- 1}
{^G \cR^{[1]}_{jk}} = \frac{1}{3} \theta_{j - 1, k}
\Bigg\{
\frac{(k + 1)(k + 3) \mp 2}{k + 1}
+ (k + 2)(k + 3) {\mit\Psi}_{jk}
+ (- 1)^{j + k}(k + 2)(k + 3) {\mit\Xi}_{jk}
\Bigg\} , \nonumber
\end{equation}
with $-(+)$ sign for vector (axial) channel; and for quarks
\begin{equation}
\label{QuarkR}
\left( \frac{\alpha_s}{2 \pi} N_c \right)^{- 1}
{^Q \cR^{[1]}_{jk}} = 2 \frac{\theta_{j - 1, k}}{(k + 1)}
\Bigg\{
(- 1)^{j + k} (k + 2) - 1
- (- 1)^{j + k} (j + 1)(j + 2) {\mit\Psi}_{jk}
- (j + 1)(j + 2) {\mit\Sigma}_{jk}
\Bigg\} . \nonumber
\end{equation}
To simplify the presentation we have introduced the matrices, depending
only on logarithmic derivative of the Euler integral $\psi (x) =
\frac{d}{dx} \ln {\mit\Gamma} (x)$, via
\begin{eqnarray}
{\mit\Psi}_{jk}\!\!\!&=&\!\!\!
\psi \left( \frac{j + k + 4}{2} \right)
- \psi \left( j + k + 4 \right)
- \psi \left( \frac{j - k}{2} \right)
+ \psi \left( j - k \right) , \nonumber\\
{\mit\Xi}_{jk}\!\!\!&=&\!\!\!
\psi \left( j + k + 4 \right)
+ \psi \left( j - k \right)
- \psi \left( k + 4 \right)
- \psi \left( k + 1 \right) , \nonumber\\
{\mit\Sigma}_{jk}\!\!\!&=&\!\!\!
\psi \left( j + k + 4 \right)
+ \psi \left( j - k \right)
- 2 \psi \left( j + 2 \right) .
\end{eqnarray}
The conformal moments $\cR_{jk}$ of the kernels $\cR (x, y)$  are related
to those in Eq.\ (\ref{DeltaAnomaly}) by a normalization factor and,
therefore, the superconformal anomaly is
\begin{equation}
\label{Deltas}
{^a\!{\mit\Delta}}_{jk} = -\frac{2}{\varrho_k}
\left( {^Q \omega^a_j} \ {^Q {\mit\cR}}_{jk}
+ {^G \omega^a_j} \ {^G {\mit\cR}}_{jk} \right)
\qquad\mbox{with}\qquad
a = 1, 2.
\end{equation}
We should note that for even and odd $j-k$  the $\psi$-functions with
argument depending on both $j$ and $k$ enter only in the particular
combination
\begin{eqnarray*}
A_{jk} = \psi\left( \frac{j + k + 4}{2} \right)
- \psi\left( \frac{j - k}{2} \right)
+ 2 \psi ( j - k ) - \psi ( j + 2 ) - \psi(1),
\end{eqnarray*}
which arises in all special conformal anomalies $\gamma^c$, too
\cite{BelMul98b}.

The final step is to insert our findings for the superconformal anomalies
given in Eqs.\ (\ref{GluonR}-\ref{Deltas}) and our results for the eight
special conformal anomaly matrices from Ref.\ \cite{BelMul98b}, rotated to
${^{ab}\!{\mit\Gamma}^c}$-basis by means of (\ref{abTOQGrotation}), into
the four conformal constraints (\ref{ConstraintFirst}-\ref{ConstraintLast}).
Indeed, we find that all of them are identically fulfilled. Thus, the
universality of the special conformal anomaly matrix for even and odd 
parity sectors arises from the form of superconformal anomaly in 
$\cN = 1$ Yang-Mills theory.

\section{Conclusions.}

The use of $\cN = 1$ supersymmetry allows to find non-trivial relations
for both scale and special conformal anomalies of QCD conformal operators.
In the former case, the constraints, at leading order in dimensional
regularization scheme as well as to all orders in a supersymmetry preserving
scheme, involve only anomalous dimensions. There are six constraints for
the diagonal and four for the non-diagonal elements of the four entries in
the anomalous dimension matrices in the parity even and odd case. Although,
in the supersymmetric limit the colour factors have to be identified, the
constraints have predictive power also beyond the leading order
approximation. As it has been observed in Ref.\ \cite{BelMueSch98} there
is a subtlety in finding the supersymmetry preserving regularization scheme
in the non-forward case. Standard dimensional reduction \cite{Sie79} does
not serve its purpose. However, an existence of a supersymmetry preserving
scheme was proven by the ability to fulfill the constraints by a
multiplicative renormalization of anomalous dimensions.

There exist four constraints for the special conformal anomalies. They
contain new ingredients, ${\mit\Delta}$, due to the superconformal
symmetry breaking by the trace anomaly, $\cA$, in the spinor current. The
four symmetry violating entries arise as a counterterm in the product of
$\cA$ and conformal operators. Thus, there is no predictive power in these
constraints, however, they serve for a consistency check of the special
conformal anomalies.

Let us mention for completeness, the situation with the two further 
leading twist-two conformal operators appearing in the definition of the
so-called transversity distributions. It is well known that these quark 
and gluon operators do not mix with each other under renormalization due 
to different spin representations w.r.t.\ Lorentz group. Their forward 
anomalous dimensions are related in the supersymmetric limit by one
constraint. However, unfortunately, there exist no constraints for the 
non-diagonal elements of conformal anomalies that do not involve anomalous 
contributions of fermionic operators \cite{BelMulDIS00}.

We have evaluated the superconformal anomalies in one-loop approximation
and found that our relations are indeed fulfilled with the known results 
for the special conformal anomalies from Refs.\ \cite{Mue94,BelMul98b}. In
other words, we can now reconstruct the special conformal anomalies of
the gluon-gluon and gluon-quark sector from those in the remaining two
channels, provided the former two would not be already known. Furthermore, 
we know that the latter give us the rotation matrix to the conformal 
scheme in which the conformal operator product expansion of two 
electromagnetic currents is valid. Since the Wilson coefficients of this
expansion in both of these remaining channels are fixed by the ones of 
the forward kinematics and coincide with the explicit calculation in the 
minimal subtraction scheme \cite{ManPilSteVanWei97JiOsb98}, rotated to the 
conformal scheme \cite{Mue97aBelMue97a}, we have a complete consistency 
check of all special conformal anomaly matrices evaluated at one-loop 
level. Since these conformal anomalies induce, together with terms 
proportional to the $\beta$ function, the non-diagonal elements of the 
anomalous dimension matrix at two-loop level, we have also a complete but 
indirect check on their correctness from the superconstraints 
(\ref{SUSYrelations}). Of course, in the flavour non-singlet sector
the anomalous dimensions arising from our prediction coincide with
the conformal moments of evolution kernel calculated at two-loop level
\cite{Sam84DitRad84MikRad85}. 

Altogether, we have a complete list of consistency checks for the field
theoretical treatment of conformal anomalies, their evaluation at one-loop
order and next-to-leading predictions arising from their use. This also
supports our results for the transversity sector \cite{BelFreMue00}, which
has been treated in the same manner.

\vspace{0.5cm}

A.B. would like to thank A. Sch\"afer for the hospitality extended to him
at the Institut f\"ur Theoretische Physik, Universit\"at Regensburg where
a final part of this work has been done. This work was supported by
Alexander von Humboldt Foundation, in part by National Science Foundation,
under grant PHY9722101 and Graduiertenkolleg Erlangen-Regensburg (A.B.),
and by DFG and BMBF (D.M.).

\appendix

\setcounter{section}{0}
\setcounter{equation}{0}
\renewcommand{\theequation}{\Alph{section}.\arabic{equation}}

\section{Evaluation of the fermionic $b$-matrix.}
\label{matrixbF}

The fermionic $b$-matrix is defined by  Eq.\ (\ref{FermiBmatrix}).
In order to evaluate the matrix it proves convenient to make the
Fourier transform on the fields, i.e.\ $\phi (x) = \int d^d k \,
{\rm e}^{- k \cdot x} \widetilde \phi (k)$ and introduce a set of
new variable of the integrand: $Y \equiv k_{2+} + k_{1+}$ and
$X = \frac{k_{2+} - k_{1+}}{k_{2+} + k_{1+}}$, so  we get
\begin{eqnarray}
\int d^d x \, 2 x_- \psi \frac{\delta}{\delta\psi} \ \cV_{jl}
\!\!\!&=&\!\!\! 2 i \, \varrho_j
\int d^d k_1 \, d^d k_2 \, Y^{l - 1}
\left\{ \hat {\cal L} \, P_j^{(2, 1)} (X) \right\}
{\widetilde G}_{+\mu} (k_1) \gamma_\mu^\perp {\widetilde\psi} (k_2)
\nonumber\\
&=&\!\!\! 2 i \sum_{k = 0}^{j} \frac{\varrho_j}{\varrho_k}
B_{jk} (2, 1) \cV_{kl-1},
\end{eqnarray}
with differential operator $\hat {\cal L} \equiv l + (1 - X)
\frac{d}{d X}$ and $B_{jk}$ being defined by the integral
\begin{equation}
B_{jk} (\alpha, \beta)
= \int_{-1}^{1} dX \frac{w (X | \alpha, \beta)}{n_k (\alpha, \beta)}
P_k^{(\alpha, \beta)} (X) \hat {\cal L} \, P_j^{(\alpha, \beta)} (X) .
\end{equation}
The weight and normalization factors are given by standard equations
\begin{eqnarray*}
w (X | \alpha, \beta) \!\!\!&=&\!\!\! (1 - X)^\alpha (1 + X)^\beta ,\\
n_k (\alpha, \beta) \!\!\!&=&\!\!\! 2^{\alpha + \beta + 1}
\frac{{\mit\Gamma} (k + \alpha + 1) {\mit\Gamma} (k + \beta + 1)}{
(2k + \alpha + \beta + 1) {\mit\Gamma} (k + 1)
{\mit\Gamma} (k + \alpha  + \beta + 1)} .
\end{eqnarray*}
Using Rodriga's formula for Jacobi polynomials and integrating $k$ (and $k +
1$) times by parts we come to integrals which can be easily evaluated by
means of the result
\begin{equation}
\int_{-1}^{1} dX \, w (X | \alpha, \gamma) P_j^{(\alpha, \beta)} (X)
= (- 1)^j 2^{\alpha + \gamma + 1}
\frac{{\mit\Gamma} (\alpha + j + 1) {\mit\Gamma} (\beta - \gamma + j)
{\mit\Gamma} (\gamma + 1)}{{\mit\Gamma} (j + 1) {\mit\Gamma}
(\beta - \gamma) {\mit\Gamma} (\alpha + \gamma + j + 2)} .
\end{equation}
So that we finally obtain
\begin{equation}
B_{jk} (\alpha, \beta)
= (l - k) \delta_{jk} - \theta_{j,k + 1} (-1)^{j - k}
(2k + \alpha + \beta + 1)
\frac{{\mit\Gamma} (j + \alpha + 1) {\mit\Gamma} (k + \alpha + \beta + 1)}{
{\mit\Gamma} (k + \alpha + 1) {\mit\Gamma} (j + \alpha + \beta + 1)} .
\end{equation}
Identifying $\alpha = 2$ and $\beta = 1$ we get the result
\begin{equation}
\label{bFmatr}
b_{jk} (F) \equiv \frac{\varrho_j}{\varrho_k}
\left\{ (l - k) \delta_{jk}
- 2 (- 1)^{j - k} \frac{(k + 2)(k + 3)}{(j + 3)} \theta_{j - 1, k}
\right\} ,
\end{equation}
where $\theta_{jk} = \{1,\ \mbox{if}\ j \geq k;\ 0,\ \mbox{if}\ j < k \}$.

\section{Renormalization of the operator product $[\cA][\cO]$.}
\label{Diagram-by-diagram}

\setcounter{equation}{0}

For the calculation of $Z$-matrices one uses in the light-ray formalism
the usual momentum space
Feynman rules and ${\rm Vertex} \times \exp \left( - i \kappa_1 k_{1+} - i
\kappa_2 k_{2+} \right)$ for the non-local operator where `Vertex' stands
for Dirac or Lorentz tensor. Introducing the Feynman parameters $x$, $y$ for
the propagators we reduce Feynman integrals to the form
\begin{equation}
\cJ \ast \cR (x, y) \times \mbox{Vertex} ,
\ \mbox{with}\
\cJ = \int_{0}^{1} dx \int_{0}^{1} dy \, \theta (1 - x - y)
e^{- i k_{1+} ( (1-x) \kappa_1 + x \kappa_2 )
- i k_{2+} ( y \kappa_1 + (1-y) \kappa_2 )} ,
\end{equation}
where the exponential corresponds to the Fourier transform of the
coordinate dependence of the string operator `after evolution', $\cO
((1-x) \kappa_1 + x \kappa_2, y \kappa_1 + (1-y) \kappa_2 ))$.
To calculate the divergent part of the operator product
$[\cA][\cO]$, we have in addition to take the Feynman rule for the anomaly
$[\cA]$ defined in Eq.\ (\ref{defA}).

Our calculation has been performed with arbitrary gauge fixing parameter
$\xi$, which has canceled in the sum of diagrams. Therefore, we present the
results corresponding to the definition (\ref{OAnonLocEE}) for the separate
contributions in $\xi = 1$ gauge.

In the quark parity even and odd sectors we get on the diagram-by-diagram
basis from Fig.\ \ref{RAquarks}
\begin{eqnarray}
- k_{2+} {^Q \cR_{(a)}}
\!\!\!&=&\!\!\! \frac{1}{2} ,
\nonumber\\
- k_{2+} {^Q \cR_{(b)}}
\!\!\!&=&\!\!\! - \frac{1}{2} (1 - y) \delta (x) ,
\nonumber\\
- k_{2+} {^Q \cR_{(c)}}
\!\!\!&=&\!\!\! \pm \frac{x}{2} \delta (1 - x - y) ,
\nonumber\\
- k_{2+} {^Q \cR_{(d)}}
\!\!\!&=&\!\!\! \frac{y}{2} \delta (1 - x - y)
- \frac{1}{2} \left( 1 + \frac{y}{2} \right) \delta (x) - \frac{1}{2} ,
\nonumber\\
- k_{2+} {^Q \cR_{(e)}}
\!\!\!&=&\!\!\! \left[ \frac{1}{y} \right]_+ \delta (x) - \delta (x) ,
\nonumber\\
- k_{2+} {^Q \cR_{(f)}}
\!\!\!&=&\!\!\! - \left[ \frac{1}{y} \right]_+ \delta (x)
+ \frac{1}{2} (3 - y) \delta (x) ,
\nonumber\\
- k_{2+} {^Q \cR_{(g)}}
\!\!\!&=&\!\!\! - \left[ \frac{1}{y} \right]_+ \delta (x)
+ \frac{1}{2} \left( 1 + \frac{y}{2} \right) \delta (x) ,
\label{DbyDquarks}
\end{eqnarray}
where $+$ ($-$) sign in the contribution of the diagram (c) stands for even
(odd) case. In these results we dropped the contributions of the type ${\rm
const} \cdot \delta (x) \delta (y)$, since they do not enter into the physical
part of the constraints (\ref{ConstraintFirst}-\ref{ConstraintLast}),
namely for $k < j$. In the sum of Eqs.\ (\ref{DbyDquarks}) the term
$\ft12 \cdot \delta (1 - x - y)$ for vector case and $\ft12 ( 1 - 2x )
\cdot \delta (1 - x - y)$ for the axial one cancels with the $\kappa_1
\leftrightarrow \kappa_2$ contribution (`$+$' and `$-$' sign,
respectively) in Eq.\ (\ref{OAnonLocEE}) and we get the result in Eq.\
(\ref{RAVkernelsQUARK}).

The gluon case is calculated from the graphs in Fig.\ \ref{RAgluons}
and reads for vector channel
\begin{eqnarray}
- k_{2+} {^G \cR^V_{(a| \Phi)}}
\!\!\!&=&\!\!\! k_{2+} \delta (y)
\left\{ \left[ \frac{2}{x} \right]_+ - 2 + 2 \delta (x) \right\},
\nonumber\\
- k_{2+} {^G \cR^V_{(a| NA)}}
\!\!\!&=&\!\!\! \delta (y)
\left\{
\frac{3}{2} (1 - x^2) k_{1+} + \frac{i}{2} \kappa x (1 - x)^2 k_{1+}^2
- \frac{3}{2} \delta (x) k_{2+}
\right\},
\nonumber\\
- k_{2+} {^G \cR^V_{(b)}}
\!\!\!&=&\!\!\!
\left( 2 - \frac{3}{2} y \right) k_{2+}
- \frac{3}{2} (1 - x) k_{1+}
+ \frac{i}{2} \kappa \left( (1 - y)k_{2+} + x k_{1+} \right)
\left( y k_{2+} + (1 - x) k_{1+} \right) \nonumber\\
&+&\!\!\! \frac{1}{2} (1 - x) \delta (y)
\left\{
(1 - 3 x) k_{1+} - 2 k_{2+}
- i \kappa (1 - x) k_{1+} \left( x k_{1+} + k_{2+} \right)
\right\} ,
\nonumber\\
- k_{2+} {^G \cR^V_{(c)}}
\!\!\!&=&\!\!\! y k_{2+} - (1 + x) k_{1+} ,
\nonumber\\
- k_{2+} {^G \cR^V_{(d)}}
\!\!\!&=&\!\!\!
\left( 2 - \frac{3}{2} y \right) k_{2+}
- \frac{3}{2} (1 - x) k_{1+}
+ \frac{i}{2} \kappa \left( (1 - y) k_{2+} + x k_{1+} \right)
\left( y k_{2+} + (1 - x) k_{1+} \right) ,
\nonumber\\
- k_{2+} {^G \cR^V_{(e)}}
\!\!\!&=&\!\!\!
\frac{1}{2} (3 - x) k_{1+} - \left( 1 - \frac{y}{2} \right) k_{2+}
- \frac{i}{2} \kappa \left( (1 - y) k_{2+} + x k_{1+} \right)
\left( y k_{2+} + (1 - x) k_{1+} \right) ,
\nonumber\\
- k_{2+} {^G \cR^V_{(f| \Phi)}}
\!\!\!&=&\!\!\! k_{2+} \delta (y)
\left\{ - \left[ \frac{1}{x} \right]_+ + 1 - \delta (x) \right\} ,
\nonumber\\
- k_{2+} {^G \cR^V_{(f| NA)}}
\!\!\!&=&\!\!\! \delta (y)
\left\{ - (1 - x) k_{1+} + \delta (x) k_{2+} \right\} ,
\nonumber\\
- k_{2+} {^G \cR^V_{(g)}}
\!\!\!&=&\!\!\!
 k_{2+} {^G \cR^V_{(d)}}  ,
\label{DbyDgluonsV}
\end{eqnarray}
with $\kappa \equiv \kappa_2 - \kappa_1$. The subscript $\Phi$ ($NA$)
stands for the contributions originating from the expansion of the
path ordered exponential (non-abelian part of the field strength tensor).
The axial case differs from the previous one only in the contributions
of diagrams (b), (c) and (e) which are
\begin{eqnarray}
- k_{2+} {^G \cR^A_{(b)}}
\!\!\!&=&\!\!\!
\left( 2 - \frac{7}{2} y \right) k_{2+}
- \left( \frac{3}{2} - \frac{7}{2} y \right) k_{1+}
+ \frac{i}{2} \kappa \left( (1 - y) k_{2+} + x k_{1+} \right)
\left( y k_{2+} + (1 - x) k_{1+} \right) \nonumber\\
&+&\!\!\!
\frac{1 - x}{2} \delta (y)
\left\{
(1 - 3 x) k_{1+} - 2 k_{2+} - i \kappa x (1 - x) k_{1+}^2
- i \kappa (1 - x) k_{1+} k_{2+}
\right\} ,
\nonumber\\
- k_{2+} {^G \cR^A_{(c)}}
\!\!\!&=&\!\!\!
- y k_{2+} - (1 - x) k_{1+} ,
\\
- k_{2+} {^G \cR^A_{(e)}}
\!\!\!&=&\!\!\!
\left( \frac{3}{2} - \frac{5}{2} y \right) k_{1+}
- \left( 1 - \frac{5}{2} y \right) k_{2+}
- \frac{i}{2} \kappa \left( (1 - y) k_{2+} + x k_{1+} \right)
\left( y k_{2+} + (1 - x) k_{1+} \right) . \nonumber
\label{DbyDgluonsA}
\end{eqnarray}
Summing the separate terms we have to use the formula (see Ref.\
\cite{BelMul98b} for a general result)
\begin{equation}
\cJ \ast \left\{
k_{1+} \left[ 1 - (1-x) \delta(y) \right]
+
k_{2+} \left[ 1 - (1-y) \delta(x) \right]
\right\} = 0 ,
\end{equation}
to reduce the result to its final form (\ref{RAVkernelsGLUON}).
Then we use the equation $k_{2+} \widetilde{B}_\mu^a (k_2) = i
\widetilde{G}_{+\mu}^a (k_2)$, valid to leading order in the coupling,
to reconstruct the field strength from the potential.

\section{Conformal moments of regularized kernels.}
\label{Moments}

\setcounter{equation}{0}

The conformal moments of the transition kernels derived in the body
of the text are defined according to
\begin{equation}
\cR^i_{jk} \equiv \int_{-1}^{1} dt \frac{w (2,1| t)}{n_k (2,1)}
P^{(2, 1)}_k (t) \int_{0}^{1} dx \int_{0}^{\bar x} \cR (x, y)
C^i_j \Big( [1 - x - y]t - x + y \Big) ,
\end{equation}
with the weight $w (2,1| t) \equiv (1 - t)^2 (1 + t)$ and normalization
$n_k (2,1) \equiv 8 \frac{(k + 1)}{(k + 2)(k + 3)}$. We have for quarks
$C^Q_j = C^{3/2}_j$ and $C^G_j = C^{5/2}_{j - 1}$ for gluons. From this
equation it is straightforward to evaluate the moments of all parts of
the kernels (along the line of Ref.\ \cite{BelMul98b}) except of the
ones with $+$-prescription since in this case we obtain the result in
terms of derivatives of hypergeometric function w.r.t.\ its indices which
is not easy to handle. In these cases we have to modify our modus
operandi and develop a more efficient machinery which leads to more
tractable expressions. It can be achieved according to the re-expansion
of the integrand making use of the orthogonality for Gegenbauer
polynomials. To be more specific let us consider the gluon kernel
$[1/x]_+ \delta (y)$ which in the momentum fraction formalism translates
into $[\theta(t - t')/(t - t')]_+$. We use the following regularization
of singular distributions
\begin{equation}
\label{PlusPrescription}
\int_{0}^{1} \frac{dx}{[1 - x]_+} \phi (x)
\equiv \int_{0}^{1} \frac{dx}{(1 - x)^{1 - \varepsilon}}
[\phi(x) - \phi(1)].
\end{equation}
Then using the representation of the Gegenbauer polynomial in terms of
hypergeometric function ${_2F_1}$ and using Rodriga's formula for
Jacobi polynomials we integrate $k$ times by parts to get
\begin{eqnarray}
\label{MomPlusDistr}
M^G_{jk} \equiv
\left\{ \left[ \frac{\theta(t - t')}{(t - t')} \right]_+ \right\}_{jk}
\!\!\!&=&\!\!\! (- 1)^{j + k} \frac{(k + 2)(k + 3)}{6 \varepsilon}
\nonumber\\
&-&\!\!\! (- 1)^j \frac{(k + 2)(k + 3)\Gamma (j + 4)}{12
\Gamma (j) \Gamma (k + 2)}
\int_{0}^{1} dx x^k (1 - x)^{\varepsilon - 1} \int_{0}^{1} dy
y^{k + 1} (1 - y)^{k + 2} \nonumber\\
&&\qquad\qquad\qquad\times\frac{d^k}{d (xy)^k}
{_2 F_1}
\left( \left. { - j + 1 , j + 4 \atop 3 }
\right| xy \right) .
\end{eqnarray}
The first term on the r.h.s.\ originates from the $\phi (1)$ contribution in
Eq.\ (\ref{PlusPrescription}). The simplicity of the consequent analysis
depends on the handling of derivatives acting on ${_2F_1}$. If we merely
differentiate it $k$ times as it appears and perform $y$ integration this
gives $_3 F_2$. Finally, after $x$ integration Eq.\ (\ref{MomPlusDistr})
will be proportional to the derivative of $_4 F_3$ w.r.t.\ a low index [see
later Eq.\ (\ref{F43derivative})]. Fortunately, it is possible to avoid this
if one notices that $k - 1$ $\varepsilon$-differentiations and $y$
integration leads to a $_2 F_1$ function with shifted indices. After the
last integration is done one ends up with an expression with $\varepsilon$
derivative acting on $_3 F_2$. In order to achieve this let us re-expand the
action of $k^{\rm th}$ derivative $d/d (xy)$ on $_2 F_1$ w.r.t.\ the
complete set of polynomials $C^{k + 3/2}_{l} (2x - 1)$
\begin{equation}
\frac{d}{dx}
{_2 F_1}
\left( \left. { - j + k , j + k + 3 \atop k + 2 }
\right| x \right)
= \sum_{l = 0}^{\infty} c_{j - k, l} \,
{_2 F_1}
\left( \left. { - l , l + 2k + 3 \atop k + 2 }
\right| x \right) ,
\end{equation}
with expansion coefficients
\begin{equation}
c_{j - k, l} = - [1 - (- 1)^{j - k - l}] \theta_{j - k, l + 1}
(2l + 2k + 3) \frac{\Gamma (l + 2k + 3)}{\Gamma (l + 1)}
\frac{\Gamma (j - k + 1)}{\Gamma (j + k + 3)}
\end{equation}
easily obtained from the orthogonality of polynomials. Then we get
\begin{eqnarray}
\label{SumC}
&&M^G_{jk}
= \frac{(- 1)^{j + k}}{12}
\frac{(k + 2)^2 (k + 3)\Gamma (j + k + 3)}{\Gamma (2k + 5)
\Gamma (j - k + 1)}
\left\{
\psi (1) - \psi (k + 1) + \frac{\partial}{\partial\varepsilon}
\right\}_{\varepsilon = 0} \nonumber\\
&&\qquad\qquad\qquad\times\sum_{l = 0}^{\infty}
c_{j - k, l} \,
{_3 F_2}
\left( \left. { - l , l + 2k + 3, k + 1 \atop 2k + 5, k + 1 + \varepsilon }
\right| 1 \right) .
\end{eqnarray}
Now the way to handle the derivative of $_3 F_2$ is rather
straightforward. First we use the fundamental identity for $_3 F_2$
($l \in {\rm I\!N}$)
\begin{equation}
{_3 F_2}
\left( \left. { - l , l + \alpha, \beta \atop \gamma, \beta + \varepsilon }
\right| 1 \right)
= \frac{\Gamma (\gamma)\Gamma (\gamma - \alpha)}{\Gamma (\gamma + l)
\Gamma (\gamma - \alpha - l)}
{_3 F_2}
\left( \left. { - l , l + \alpha, \varepsilon
\atop
1 + \alpha - \gamma, \beta + \varepsilon }
\right| 1 \right) ,
\end{equation}
where we substitute $\gamma = \alpha + 2 - \rho$ with $\rho \to 0$.
Then the expansion w.r.t.\ $\varepsilon$ is easy to construct
\begin{eqnarray}
&&{_3 F_2}
\left( \left. { - l , l + \alpha, \varepsilon
\atop
\rho - 1 , \beta + \varepsilon }
\right| 1 \right)
= 1 + \varepsilon\,
\Bigg\{
\frac{l (l + \alpha)}{\beta} \\
&&\qquad\qquad+ \Gamma (\rho - 1)
\Bigg(
1 + (- 1)^l
\left[ (l - 1)(l + \alpha + 1) + \beta \right]
\frac{\Gamma (1 + l + \alpha - \beta) \Gamma (\beta)}{
\Gamma (2 + \alpha - \beta) \Gamma (\beta + l)}
\Bigg)
\Bigg\}
+\cO (\varepsilon^2) , \nonumber
\end{eqnarray}
and together with the identity
\begin{equation}
\frac{\Gamma (\rho - 1)}{\Gamma (2 - \rho - l)}
= (- 1)^l
\frac{\Gamma (l - 1 + \rho)}{\Gamma (2 - \rho)}
\end{equation}
we find
\begin{eqnarray}
\label{F32expand1}
&&{_3 F_2}
\left( \left. { - l , l + \alpha, \beta \atop \alpha + 2, \beta + \varepsilon }
\right| 1 \right)
= \frac{\Gamma (\alpha + 2)}{\Gamma (l + \alpha + 2)}
\Bigg\{
\left( 1 + \varepsilon\, \frac{l (l + \alpha)}{\beta} \right)
\left( \delta_{l, 0} + \delta_{l, 1} \right) \\
&&\qquad\qquad+ \varepsilon\,
\Gamma (l - 1)
\Bigg(
(- 1)^l + \left[ (l - 1)(l + \alpha + 1) + \beta \right]
\frac{\Gamma (1 + l + \alpha - \beta) \Gamma (\beta)}{
\Gamma (2 + \alpha - \beta) \Gamma (\beta + l)}
\Bigg) \theta_{l, 2}
+\cO (\varepsilon^2)
\Bigg\} . \nonumber
\end{eqnarray}
Using the results we have just derived, we perform in a last step the
summation in Eq.\ (\ref{SumC}) according to the formula
\begin{eqnarray}
&&\left. \frac{\partial}{\partial\varepsilon} \right|_{\varepsilon = 0}
\sum_{l = 0}^{\infty}
c_{j - k, l} \,
{_3 F_2}
\left( \left. { - l , l + 2k + 3, k + 1 \atop 2k + 5, k + 1 + \varepsilon }
\right| 1 \right)
= \frac{1}{(k + 1)(k + 2)}
- \frac{(j - k)(j + k + 3)}{(k + 1)(k + 2)(k + 3)} \nonumber\\
&&\qquad\qquad- \frac{1}{k + 2}
\Bigg\{
(- 1)^{j + k} \psi \left( \frac{j + k + 4}{2} \right)
+ [1 - (- 1)^{j + k}] \psi (j + k + 4)
- (- 1)^{j + k} \psi \left( \frac{j - k}{2} \right) \nonumber\\
&&\qquad\qquad+ [1 + (- 1)^{j + k}] \psi (j - k)
- \psi (k + 4) - \psi (1)
\Bigg\} .
\end{eqnarray}
As a by-product we verify the following formula for the derivative of
$_4 F_3$, which is difficult to derive by other means
\begin{eqnarray}
\label{F43derivative}
&&\frac{\Gamma (j + k + 4)}{\Gamma (j - k)\Gamma (2k + 5)}
\left. \frac{\partial}{\partial\varepsilon} \right|_{\varepsilon = 0}
{_4 F_3}
\left( \left. { - j + k + 1 , j + k + 4, k + 2, k + 1
\atop
2 k + 5, k + 3, k + 1 + \varepsilon }
\right| 1 \right) \nonumber\\
&&\qquad\qquad=
\frac{(j - k)(j + k + 3)}{(k + 1)(k + 3)} - \frac{1}{k + 1}
+
(- 1)^{j + k} \psi \left( \frac{j + k + 4}{2} \right)
+ [1 - (- 1)^{j + k}] \psi (j + k + 4) \nonumber\\
&&\qquad\qquad- (- 1)^{j + k} \psi \left( \frac{j - k}{2} \right)
+ [1 + (- 1)^{j + k}] \psi (j - k)
- \psi (k + 4) - \psi (1) .
\end{eqnarray}

Slightly different procedure holds for quarks. In this case due to the
presence of the momentum fraction $k_{2 +} = \frac{1 - t}{2}
(k_1 + k_2)_+$ we re-expand the integrand, modified by adding a constant,
\begin{eqnarray}
M^Q_{jk} \!\!\!&\equiv&\!\!\!
\left\{ \left[ \frac{2}{1 - t}
\frac{\theta(t' - t)}{(t' - t)} \right]_+ \right\}_{jk}
= (- 1)^k \frac{(k + 2)(k + 3)(j + 1)(j + 2)}{k + 1} \\
&\times&\!\!\!
\int_{0}^{1} y^2 (1 - y) P^{(1, 2)}_k (2 y - 1)
\int_{0}^{1} dx \, \left[ \frac{1}{x} \right]_+
\frac{1}{y}
\left\{
{_2 F_1}
\left( \left. { - j , j + 3 \atop 2 }
\right| xy \right)
-
{_2 F_1}
\left( \left. { - j , j + 3 \atop 2 }
\right| 0 \right)
\right\} , \nonumber
\end{eqnarray}
in the following series
\begin{equation}
\frac{1}{x}
\left\{
{_2 F_1}
\left( \left. { - j , j + 3 \atop 2 }
\right| x \right)
-
{_2 F_1}
\left( \left. { - j , j + 3 \atop 2 }
\right| 0 \right)
\right\}
= \sum_{k = 0}^{\infty} d_{jk} \,
{_2 F_1}
\left( \left. { - k , k + 3 \atop 2 }
\right| x \right)
\end{equation}
with the expansion coefficients
\begin{equation}
d_{jk} = - \theta_{j - 1, k} \frac{3 + 2k}{(j + 1)(j + 2)}
(j - k)(j + k + 3) .
\end{equation}
Consequent integration and expansion in $\varepsilon$ requires
Eq.\ (\ref{F32expand1}) as well as the following result
\begin{eqnarray}
&&{_3 F_2}
\left( \left. { - l , l + \alpha , \beta
\atop 1 + \alpha , \beta + \varepsilon }
\right| 1 \right) \\
&&\qquad\qquad= \frac{\Gamma (1 + \alpha)}{\Gamma (1 + \alpha + l)}
\left\{
\delta_{l,0}
+
\varepsilon \, \Gamma (l)
\left(
\frac{\Gamma (1 + \alpha - \beta + l) \Gamma (\beta)}{
\Gamma (1 + \alpha - \beta) \Gamma (\beta + l)}
- (- 1)^l
\right) \theta_{l, 1}
\right\} + {\cal O} (\varepsilon^2) , \nonumber
\end{eqnarray}
which can be deduced using the same recipe as presented above.
Final summation gives us the result in Eq.\ (\ref{QuarkR}).

\end{document}